\documentclass[twocolumn,aps,prb,groupedaddress,superscriptaddress,amsmath,floatfix,amssymb,showpacs,noeprint,english]{revtex4-2}

\usepackage[T1]{fontenc}
\usepackage{graphicx}
\usepackage{bm}
\usepackage{dcolumn}
\usepackage{float}
\usepackage{capt-of}
\usepackage{mathrsfs}
\usepackage{amsbsy}
\usepackage{amstext}
\usepackage{amsmath,amsthm,amsfonts,amssymb,bbold}
\DeclareMathOperator{\Tr}{Tr}
\usepackage{setspace}
\usepackage{esint}
\usepackage{ragged2e}
\usepackage{placeins}


\usepackage{booktabs}
\usepackage{threeparttable}
\usepackage{multirow}
\usepackage{float}
\usepackage{placeins}

\usepackage{algorithm}
\usepackage{algpseudocode}
\usepackage{comment}

\usepackage{tikz}
\usepackage{quantikz}

\usepackage{xcolor,colortbl}
\usepackage[table]{xcolor} % Permite colorir linhas e colunas
\usepackage{array}         % Necessário para injetar comandos nas colunas (ex: >{\columncolor})
\usepackage{makecell}      % Facilita quebras de linha nas células

\usepackage[english]{babel}

\usepackage{subcaption}

\usepackage{url}
\usepackage[hidelinks,hypertexnames=true]{hyperref}

\begin{document}

\title{Generalized two-qubit Hamiltonian for Projective Quantum Feature Maps}

\author{Rafael~Sim\~oes~do~Carmo}
\affiliation{Faculty of Sciences, UNESP - S\~ao Paulo State University, 17033-360 Bauru-SP, Brazil}

\author{Edson Amaro Junior}
\affiliation{Hospital Israelita Albert Einstein, 05652-900 S\~ao Paulo-SP, Brazil}

\author{Felipe F. Fanchini}
\affiliation{Faculty of Sciences, UNESP - S\~ao Paulo State University, 17033-360 Bauru-SP, Brazil}
\affiliation{Hospital Israelita Albert Einstein, 05652-900 S\~ao Paulo-SP, Brazil}

\date{\today}

\begin{abstract}

Projected quantum feature maps provide a strategy for using quantum processors as feature generators for classical machine-learning models. Building on counterdiabatic Ising-glass and one-dimensional Heisenberg PQFMs, we introduce a generalized two-qubit Hamiltonian-based PQFM that provides a unified way to encode classical features through local Pauli fields and pairwise two-qubit Pauli interactions. This construction allows distinct classical variables to be embedded along different Pauli axes of the same qubit, increasing the information density of shallow circuits while remaining compatible with hardware constraints. We develop and implement these methods in \texttt{pqfmlib}, a publicly available Python library for constructing, executing, and benchmarking Hamiltonian-based PQFMs.We then benchmark the generalized Hamiltonian PQFMs against reference PQFMs on four biomedical classification datasets under a nested cross-validation protocol with paired statistical tests. Quantum features are generated using both IBM quantum processors with up to 156 qubits and statevector simulations. 
Our results show that the generalized two-qubit Hamiltonian family provides the most consistent pattern of statistically supported gains over matched classical baselines, although the performance of all methods depends on the dataset, encoding strategy, measured observables, and hardware conditions. These findings support generalized Hamiltonian PQFMs as a promising route toward near-term quantum utility.

\end{abstract}

\maketitle

\section{Introduction}
Quantum computers are a promising platform for addressing computational problems that are difficult for classical machines \cite{boixo2018characterizing,arute2019quantum}. In the context of learning algorithms, one of their main promises is the possibility of improving predictive performance in tasks such as classification \cite{rebentrost2014qsvm, farhi2018classification, schuld2019feature, havlicek2019supervised, huang2021power}.  However, given the limitations of current quantum processors, which are strongly constrained by noise and restricted circuit depth, it is unlikely that near-term devices will provide broad, general-purpose advantages for machine learning \cite{preskill2018quantum, schuld2022quantum, bharti2022nisqalgorithms}. Even so, one promising use of quantum computers is quantum feature generation, where a quantum processor is used not to train a complete model, but to transform classical inputs into new features for downstream classical classifiers \cite{huang2021power}. Projected quantum feature maps (PQFMs) follow precisely this strategy and therefore provide a practical route for exploring whether current quantum devices can contribute useful information to classification workflows \cite{ferenczi2025credit, ciceri2025enhanced, simen2025quenched, simen2025digitized, zhang2026quantum}.

In a PQFM, classical data are encoded into a quantum state through a data-dependent circuit, and a finite set of expectation values is then measured to produce classical features. These quantum-generated features can be stored and used with standard classical classifiers, either alone or in combination with the original input variables \cite{ferenczi2025credit, ciceri2025enhanced, simen2025quenched, simen2025digitized, zhang2026quantum}. Recent studies have shown that Hamiltonian-based PQFMs can provide useful representations in applied learning problems. Heisenberg-inspired PQFMs have been investigated in financial and credit-risk applications \cite{ferenczi2025credit}, where quantum-generated features were used as offline inputs for classical predictive models. In parallel, Ising-glass-based quantum feature extraction, including digitized counterdiabatic constructions, has been explored in Molecular Toxicity \cite{ simen2025digitized}, medical classification, and satellite-image tasks \cite{zhang2026quantum}. Taken together, these works suggest that projected quantum features can be useful in selected regimes.

Despite these advances, previous implementations have focused on a limited set of Hamiltonian families \cite{ferenczi2025credit, ciceri2025enhanced, simen2025quenched, simen2025digitized, zhang2026quantum}. This restricts the representational power of the quantum embedding and obscures the role of the feature map itself. To address this limitation, we introduce a generalized two-qubit Hamiltonian PQFM, implemented in \texttt{pqfmlib} \cite{carmo_pqfmlib_2026}, a hardware-aware Python library for constructing, executing, and evaluating Hamiltonian-based PQFMs. This construction supports diagonal, cross-axis, and full two-local Pauli interaction structures within the same formalism, while also implementing the counterdiabatic Ising-glass and Heisenberg maps as two reference families motivated by previous studies.

A central novelty of this generalized construction is axis-resolved feature encoding, which allows different classical variables to be encoded along different Pauli axes of the same qubit. By exploiting orthogonal directions within the single-qubit Hilbert space, this approach utilizes the total volume of the Bloch sphere, thereby mitigating representational overlap between independent features. Furthermore, cross-axis interactions (such as $XY$ or $XZ$) generate much richer entanglement dynamics than purely Ising ($ZZ$) or isotropic Heisenberg models \cite{Wiersema2024}. This multi-axis strategy maximizes information density for shallow circuits, increasing the number of encoded features per qubit while allowing pairwise statistical dependencies to be embedded through diverse two-qubit Pauli channels. We also consider shared-feature configurations, in which the same feature is encoded along multiple axes, as well as multilayer constructions for feature vectors larger than the available qubit register. To the best of our knowledge, this is the first Hamiltonian-based PQFM study to systematically explore axis-resolved encoding of distinct classical features.

In order to test PQFM properties in real-life problems, we evaluate these PQFM families on four public healthcare (medical and biomedical) classification datasets using matched nested cross-validation protocols. Despite its paramount importance to human health and society, healthcare data are known for its substantial class imbalance, presenting an interesting case for the development of quantum algorithms \cite{Roesler2026.03.04.26347634}. The experiments include executions on real IBM quantum processors with up to 156 qubits, as well as statevector simulations for complementary benchmarks. For each dataset and feature-map family, we compare models trained on classical features, quantum-generated features, and combined classical–quantum representations. Performance differences are analyzed fold by fold using 95\% confidence intervals for paired metric differences and one-sided Wilcoxon signed-rank tests \cite{wilcoxon_1, GibbonsChakraborti2010}. This non-parametric testing protocol allows us to distinguish consistent improvements from merely suggestive trends across datasets, feature-map designs, and measured observables. Our results indicate that the usefulness of projected quantum features depends on the dataset, encoding strategy, measured observables, post-processing method, and hardware conditions. Nevertheless, the analysis shows that generalized multi-axis PQFMs can provide consistent and statistically supported predictive gains over matched classical baselines in selected hardware settings at utility-relevant scales.

To present our results, this paper is organized as follows. Section II introduces the PQFM framework and presents the CD-Ising-glass, Heisenberg, and generalized two-qubit Hamiltonian maps considered in this work. Section III describes the experimental methodology, including the datasets, the quantum-hardware and simulation settings, the feature-generation protocol implemented in \texttt{pqfmlib}, the classical learning pipeline, the hybrid feature-selection strategies, and the statistical evaluation protocol. Section IV presents the benchmarking results, including real-hardware experiments and statevector simulations. Section V summarizes the main findings and discusses future directions for generalized Hamiltonian PQFMs.

\section{Projected Quantum Feature Maps}
\label{sec:pqfm}

A projected quantum feature map can be understood as a kind of feature engineering technique within a machine-learning pipeline, in which a quantum circuit is used to implement a map $\phi : \mathbb{R}^{p} \rightarrow \mathbb{R}^{q}$. The basic idea is to transform an original classical feature vector into a new classical representation by exploiting the dynamics of quantum states in the $2^{n}$-dimensional Hilbert space of an $n$-qubit system. Inspired by the projected quantum kernel framework introduced in Ref.~\cite{huang2021power}, classical data are first embedded into quantum states through a quantum feature map, as commonly done in quantum machine-learning protocols. The distinctive step is that, instead of constructing a variational quantum classifier or directly evaluating a full quantum kernel matrix, a PQFM extracts explicit classical feature vectors from the encoded quantum states through measurements of a prescribed set of observables.

Formally, let $\mathcal{X} \subset \mathbb{R}^{p}$ denote the original feature space, so that each input sample is represented by a classical feature vector $\mathbf{x}=(x_1,\ldots,x_p)^{\mathsf{T}}\in\mathcal{X}$. Let $\mathcal{H}_{Q} = (\mathbb{C}^{2})^{\otimes n}$ be the Hilbert space of the $n$-qubit quantum register, and let $\mathcal{X}' \subset \mathbb{R}^{q}$ denote the transformed feature space. A PQFM is defined as the composite map
\begin{equation}
    \phi := \phi_{\mathcal{M}} \circ \phi_{U} :
    \mathcal{X}
    \xrightarrow{\;\phi_{U}\;}
    \mathcal{H}_{Q}
    \xrightarrow{\;\phi_{\mathcal{M}}\;}
    \mathcal{X}' ,
    \label{eq:pqfm_map}
\end{equation}
where $\phi_{U}$ denotes the quantum embedding map and $\phi_{\mathcal{M}}$ denotes the measurement-based projection map \cite{ciceri2025enhanced}. The embedding map is induced by a data-dependent unitary operator $\hat{U}_{\mathbf{x}} \in \mathcal{U}(\mathcal{H}_{Q})$, which evolves a fixed fiducial state $\ket{\psi_{0}}$ into a quantum feature state parametrized by the classical input vector $\mathbf{x} \in \mathcal{X}$:
\begin{equation}
    \ket{\psi_{\mathbf{x}}}
    =
    \phi_{U}(\mathbf{x})
    =
    \hat{U}_{\mathbf{x}} \ket{\psi_{0}} .
    \label{eq:pqfm_embedding}
\end{equation}

The projection map $\phi_{\mathcal{M}}$ is specified by a measurement protocol $\mathcal{M}$ characterized by a fixed set of $q$ Hermitian observables,
\begin{equation}
    \left\{
    \hat{O}_{j}
    \right\}_{j=1}^{q},
    \qquad
    \hat{O}_{j}
    =
    \hat{O}_{j}^{\dagger}
    \in
    \mathcal{B}(\mathcal{H}_{Q}) .
\end{equation}
Each transformed feature is defined as the expectation value of one observable with respect to the encoded state,
\begin{equation}
    x'_{j}
    =
    \left\langle
    \hat{O}_{j}
    \right\rangle_{\psi_{\mathbf{x}}}
    =
    \bra{\psi_{\mathbf{x}}}
    \hat{O}_{j}
    \ket{\psi_{\mathbf{x}}},
    \qquad
    j = 1, \ldots, q ,
    \label{eq:pqfm_feature_component}
\end{equation}
In practice, these expectation values are estimated from repeated circuit executions and measurements. For simplicity, Eqs.~\eqref{eq:pqfm_embedding}--\eqref{eq:pqfm_feature_component} are written in pure-state notation. On noisy quantum hardware, the same projected features can be expressed as $x'_{j} = \Tr(\rho_{\mathbf{x}} \hat{O}_{j})$, where $\rho_{\mathbf{x}}$ denotes the effective state prepared by the device.

Therefore, both the input $\mathbf{x}$ and the output $\mathbf{x}'$ of a PQFM are classical quantities. The quantum processor is used only to implement the intermediate feature transformation. Once generated, the projected quantum features may be stored in classical memory and used directly in classical learning algorithms, concatenated with the original classical variables, or employed to construct a kernel in a subsequent classical post-processing step~\cite{ciceri2025enhanced,ferenczi2025credit}.

Relative to fidelity-based quantum kernel methods, projected quantum features can provide both practical and representational advantages. Fidelity-based kernels construct a kernel matrix from pairwise similarities between quantum states encoded from the input samples, typically through
\begin{equation}
    K^{Q}_{ij}
    =
    \kappa^{Q}(\mathbf{x}_{i},\mathbf{x}_{j})
    =
    \left| \langle \psi_{\mathbf{x}_{i}} \mid \psi_{\mathbf{x}_{j}} \rangle \right|^2,
    \qquad
    i,j=1,\ldots,N ,
    \label{eq:fidelity_quantum_kernel}
\end{equation}
where $N$ is the number of samples. Since the complete kernel matrix requires evaluating similarities for all sample pairs, its quantum-evaluation cost grows as $\mathcal{O}(N^{2})$. By contrast, for a fixed measurement protocol, a PQFM generates an explicit feature vector $\mathbf{x}'_{i}=\phi_{\mathcal{M}}(\ket{\psi_{\mathbf{x}_{i}}})$ independently for each input sample, so that the quantum feature-generation stage grows linearly with $N$ for a fixed number of measured observables. If a kernel model is subsequently desired, a kernel matrix may be constructed classically from these projected features.

Beyond this resource consideration, Huang \textit{et al.} \cite{huang2021power} showed that fidelity-based quantum kernels may become poorly suited for prediction when the encoded states occupy an effectively high-dimensional region of Hilbert space. In this regime, distinct embedded samples may become nearly orthogonal,
\begin{equation}
    \left| \langle \psi_{\mathbf{x}_{i}} \mid \psi_{\mathbf{x}_{j}} \rangle \right|^2
    \approx 0,
    \qquad
    i \neq j ,
\end{equation}
so that the associated kernel matrix approaches the identity,
\begin{equation}
    K^{Q} \approx I_{N}.
    \label{eq:quantum_kernel_identity}
\end{equation}
As a consequence, the kernel provides little meaningful information about relative similarities among different inputs and may exhibit only a small geometric distinction from suitably chosen classical models~\cite{huang2021power}. 

To overcome this limitation, projected quantum kernels were introduced to map encoded quantum states back to an approximate classical representation through measurements of physically motivated observables, such as local Pauli expectation values or reduced density-matrix information \cite{huang2021power}. This idea is the origin of PQFMs, with the key distinction that, in a PQFM, the measured quantities are used directly as explicit classical features for downstream learning models. The resulting projected representation defines a new geometry in the induced feature space and can increase the geometric separation between quantum and classical models, which is a necessary, although not sufficient, condition for predictive quantum advantage.

\subsection{Ising-Glass model using CD terms}

The Ising-glass model of quantum feature extraction was first proposed in Ref.~\cite{simen2025quenched} as a technique inspired by the quench dynamics of a quantum spin glass investigated on D-Wave quantum annealers~\cite{king2025beyond}. Although the terminology of projected quantum feature maps was introduced later in the context of Heisenberg-based maps, the Ising-glass construction already fits naturally within the PQFM paradigm. Since these dynamics in the fast coherent regime have been argued to access beyond-classical simulation regimes, the motivation was to encode data samples into such a system and exploit its limited classical simulability as a rich quantum feature map. In this case, each sample is mapped to an Ising Hamiltonian, where feature values are encoded as local fields, while classical correlations are encoded as couplings. The system is then driven through a fast, non-adiabatic evolution (using D-Wave fast annealing regime), producing a highly entangled state far from the adiabatic limit. Quantum features are finally extracted by measuring expectation values of observables $\langle Z \rangle$ on the evolved state, yielding an implicit high-dimensional representation of the input. 

In subsequent work~\cite{simen2025digitized}, the authors extended the method to IBM gate-based quantum processors by digitizing the quench evolution and incorporating counterdiabatic protocols to enable fast, non-adiabatic feature generation in circuit form. This digitized version introduced Hamiltonians with higher-order correlations, for which the authors considered up to three-body coupling terms and extracted up to 3-local quantum features, with the aim of exploring richer representational structures. The resulting Hamiltonian-based feature map is defined as
\begin{equation}
H(\mathbf{x}) =
\sum_{i=1}^{n} x_i \sigma_i^z
-
\sum_{k=2}^{K}
\sum_{S \in \mathcal{G}^{(k)}}
c_S
\prod_{i \in S} \sigma_i^z ,
\label{eq:ising_glass_hamiltonian}
\end{equation}
where $x_i$ is the $i$-th component of the input feature vector $\mathbf{x}$, and the interaction weights $c_S$ are obtained from statistical dependencies among the variables associated with the subset $S$. In practice, for a subset $S$ containing $m$ variables, $c_S$ is defined as the average pairwise mutual information over all pairs contained in $S$,
\begin{equation}
c_S =
\frac{1}{\binom{m}{2}}
\sum_{\{a,b\} \subset S}
I(x_a,x_b),
\label{eq:mutual_information_coefficients}
\end{equation}
where $I(x_a,x_b)$ denotes the pairwise mutual information between the random variables $x_a$ and $x_b$ estimated from the dataset. The hypergraph $\mathcal{G}^{(k)}$ specifies which subsets $S$ contribute to the $k$-body interaction structure, thereby determining which data-dependent correlations are embedded into the quantum Hamiltonian and, when executed on hardware, into the available coupling structure of the target quantum processor.

\begin{figure*}[!t]
\centering
\includegraphics[width=\textwidth]{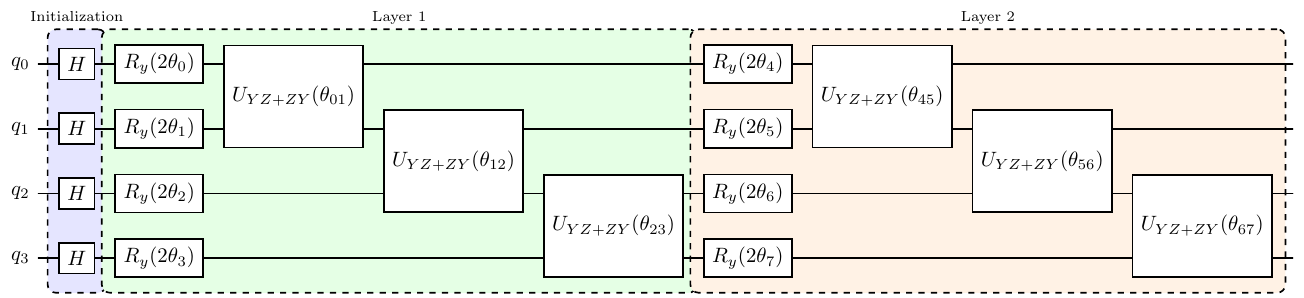}
% \captionsetup{
%     justification=justified,
%     singlelinecheck=false,
%     font=small
% }
\caption{\justifying
Circuit representation of a CD-Ising-glass with eight features mapped onto a four-qubit register. After initialization with Hadamard gates, each encoding layer applies local rotations $R_y(2\theta_i)$ followed by two-qubit interaction blocks $U_{YZ+ZY}(\theta_{ij})$. The local angles encode the one-body fields $h_i$, which are assigned from normalized classical input features, while the interaction angles encode the couplings $J_{ij}$, obtained from pairwise feature correlations, here estimated through mutual information, as defined in Eq. \ref{eq:thetas_CDIsing}. The multilayer structure enables multi-feature encoding on a fixed qubit register by assigning different feature blocks to successive encoding layers. The feature assignment follows an edge-first rule, in which highly correlated feature pairs are preferentially mapped to neighboring qubits compatible with the selected QPU topology. Through this layer-wise construction, feature vectors larger than the number of qubits can be represented while preserving the correlation-informed structure of the two-body terms within each encoding layer. The nearest-neighbor chain shown here is illustrative; in the hardware-aware implementation, the interaction pattern is determined by the processor coupling map.
}
\label{fig:circuit-4q-2layers}
\end{figure*}

To implement this PQFM on gate-based devices, the data-dependent Hamiltonian is embedded into an interpolating adiabatic Hamiltonian,
\begin{equation}
H_{\mathrm{ad}}(t;\mathbf{x}) = A(t)H_i + B(t)H(\mathbf{x}),
\end{equation}
where $H_i$ is an initial driver Hamiltonian and $A(t)$ and $B(t)$ define the annealing schedule. An approximate adiabatic gauge potential is then constructed using the nested-commutator approach \cite{claeys2019floquet}, truncated at first order,
\begin{equation}
\mathcal{A}(t;\mathbf{x})
=
i\alpha
\left[
H_{\mathrm{ad}}(t;\mathbf{x}),
\partial_t H_{\mathrm{ad}}(t;\mathbf{x})
\right],
\end{equation}
where the scalar coefficient $\alpha$ can be determined analytically, leading to a Trotterized feature-map unitary of the form
\begin{equation}
U_T(\mathbf{x})
=
\prod_{j=1}^{N_{\mathrm{steps}}}
\exp\left\{
-i\Delta t
\left[
H_{\mathrm{ad}}(j\Delta t;\mathbf{x})
+
\mathcal{A}(j\Delta t;\mathbf{x})
\right]
\right\},
\end{equation}
with $T=N_{\mathrm{steps}}\Delta t$. In the fast-driving regime considered, the authors argue that a single Trotter step is sufficient to implement the feature-generating dynamics ~\cite{simen2025digitized}. This simplification is motivated by the impulse regime, in which the counterdiabatic contribution dominates the evolution \cite{Cadavid2024counterdiabaticportfolio}. This observation also allows one to consider a reduced implementation in which the adiabatic term is neglected and the feature map is generated primarily by the counterdiabatic Hamiltonian.

In our implementation of the CD-Ising-glass map, we focus on two-body correlations and consider an adiabatic Hamiltonian of the form $H_{\mathrm{ad}}(t) = [1-\lambda(t)]H_I + \lambda(t)H_F$, where the initial transverse-field Hamiltonian is $H_I = -\sum_{i=1}^{n}\sigma_i^x$ and the final Ising Hamiltonian is
\begin{equation}
    H_F(\mathbf{x})
    =
    \sum_{i=1}^{n}
    h_i \sigma_i^z
    +
    \sum_{(i,j) \in \mathcal{G}^{(k,e)}}
    J_{ij}\,
    \sigma_i^z \sigma_j^z .
    \label{eq:cd_ising_final_hamiltonian}
\end{equation}
Here, \(h_i\) denotes the normalized value of the classical feature assigned to qubit \(i\), computed as \(h_i=(x_i-\mu_i)/\sigma_i\), with \(\mu_i\) and \(\sigma_i\) estimated from the training set. The coupling \(J_{ij}\) quantifies the statistical dependence between features \(i\) and \(j\), here estimated through mutual information. The set \(\mathcal{G}^{(k,e)}\) denotes the hardware-compatible interaction graph used for the encoding. Its edges correspond to available two-qubit connections in the selected QPU and can be weighted by calibration information, such as two-qubit gate errors. The chosen schedule function, satisfying the conditions $\lambda(0) = 0$ and $\lambda(T) = 1$, is
\begin{equation}
    \lambda(t)=\sin^{2}\!\left(\frac{\pi}{2}\,\sin^{2}\!\left(\frac{\pi t}{2T}\right)\right),
\end{equation}
which is a common choice in works dealing with counterdiabatic evolution \cite{claeys2019floquet, Cadavid2024counterdiabaticportfolio, Huerta-Ruiz_2025}. For this schedule and the adiabatic Hamiltonian defined above, the first-order nested-commutator approximation to the adiabatic gauge potential can be evaluated analytically, yielding~\cite{cadavid2023efficient}
\begin{equation}
    A_{\lambda}^{(1)}=-2\alpha_{1}\left[\sum_{i} h_{i}Y_{i}+\sum_{i<j} J_{ij}\left(Y_{i}Z_{j}+Z_{i}Y_{j}\right)\right],
    \label{eq:A_lambda}
\end{equation}
where
\begin{equation}
    \alpha_{1}=-\frac{1}{4}\,\frac{\sum_{i} h_{i}^{2}+\sum_{i<j} J_{ij}^{2}}{R(t)}\,    
    \label{eq:alpha1}
\end{equation}
with
\[
\begin{aligned}
R(t)={}&(1-\lambda)^{2}
\left(\sum_{i} h_{i}^{2}+4\sum_{i\neq j} J_{ij}^{2}\right) \\
&+\lambda^{2}\Bigg(
\sum_{i} h_{i}^{4}
+\sum_{i\neq j} J_{ij}^{4}
+6\sum_{i\neq j} h_{i}^{2}J_{ij}^{2} \\
&\qquad\qquad
+6\sum_{i<j<k}
\left(
J_{ij}^{2}J_{ik}^{2}
+J_{ij}^{2}J_{jk}^{2}
+J_{ik}^{2}J_{jk}^{2}
\right)
\Bigg).
\end{aligned}
\]
This counterdiabatic term in Eq. \ref{eq:A_lambda} is the actual Hamiltonian that we implement as a PQFM for a fiducial state prepared by applying one Hadamard gate to each qubit $\ket{\psi_0} = H\ket{0}^{\otimes n}$. Thus, for the CD-Ising-glass implementation, the feature-map unitary is approximated by a first-order Trotter product of local and two-body counterdiabatic blocks,
\begin{equation}
U_{\mathrm{CD}}(\mathbf{x})
\simeq
\prod_{s=1}^{N_{\mathrm{steps}}}
\left[
\prod_i
U_Y^{(i)}\!\left(\theta_i^{(s)}\right)
\prod_{(i,j)\in \mathcal{E}}
U_{YZ+ZY}^{(i,j)}\!\left(\theta_{ij}^{(s)}\right)
\right],
\label{eq:cd_ising_trotter_blocks}
\end{equation}
where $\mathcal{E}$ denotes the set of selected two-qubit interactions mapped to neighboring qubits of the target QPU. The elementary blocks are defined as
\begin{equation}
U_Y^{(i)}\!\left(\theta_i^{(s)}\right)
=
\exp\!\left[
i\,2\Delta t\,\alpha_1(t_s)\,
h_i\,Y_i
\right],
\label{eq:local_y_block}
\end{equation}
and
\begin{equation}
U_{YZ+ZY}^{(i,j)}\!\left(\theta_{ij}^{(s)}\right)
=
\exp\!\left[
i\,2\Delta t\,\alpha_1(t_s)\,
J_{ij}
\left(
Y_iZ_j+Z_iY_j
\right)
\right],
\label{eq:yzzy_block}
\end{equation}
with $t_s=s\Delta t$. Equivalently, the circuit parameters can be written as
\begin{equation}
\theta_i^{(s)}
=
-2\Delta t\,\alpha_1(t_s)\,h_i,
\qquad
\theta_{ij}^{(s)}
=
-2\Delta t\,\alpha_1(t_s)\,J_{ij}.
\label{eq:thetas_CDIsing}
\end{equation}
In our experiments, following the original CD-Ising-glass construction, we use $N_{\mathrm{steps}}=1$, corresponding to a single Trotter step in the impulse regime. Nevertheless, \texttt{pqfmlib} allows larger values of $N_{\mathrm{steps}}$, in which case the same sequence of local and two-body blocks is repeated with the corresponding parameters at each step.

Regarding quantum feature extraction, we use expectation values of single-qubit (one-local) and two-qubit (two-local) observables along the $Z$ axis,
\begin{equation}
x'_i
=
\langle \psi_f \vert \sigma_i^z \vert \psi_f\rangle,
\qquad
x'_{ij}
=
\langle \psi_f \vert \sigma_i^z\sigma_j^z \vert \psi_f\rangle ,
\end{equation}
where $i,j=1,\dots,n$, and $n$ denotes the number of qubits in the encoding register. High-order $Z$-correlations can also be implemented, but here we restrict the quantum feature vector to one-local and two-local observables, namely $\langle Z_i\rangle$ and $\langle Z_iZ_j\rangle$. Following the original construction, the two-local features $x'_{ij}$ are primarily extracted from pairs of neighboring qubits in the coupling topology of the selected QPU. Nevertheless, \texttt{pqfmlib} also supports the extraction of all pairwise $x'_{ij}$ features when required.

Originally, this approach enables a one-to-one encoding strategy, in which each classical feature is assigned to a single qubit. Inspired by the multilayer encoding used in the Heisenberg PQFM implementation of ~\cite{ferenczi2025credit}, \texttt{pqfmlib} extends this construction to support multi-feature encoding by increasing the number of encoding layers. The first layer is built from the feature pairs with the largest mutual information, which are mapped onto neighboring qubits according to the available QPU topology. Subsequent layers are constructed by repeating the same procedure over the remaining features, selecting the next most informative pairs and assigning them to the available interaction structure. Figure~\ref{fig:circuit-4q-2layers} illustrates this strategy in a simple setting, where eight features are encoded into a four-qubit circuit using two encoding layers. Through this layer-wise construction, feature vectors larger than the number of physical qubits can be encoded across multiple layers while preserving the correlation-informed structure of the two-body terms. When a given layer contains fewer remaining features than available encoding slots, the unused positions are filled with identity operations.

For hardware executions, \texttt{pqfmlib} uses a hardware-aware feature-assignment procedure inspired by the genetic variable-to-qubit assignment introduced in the CD-Ising-glass framework~\cite{simen2025digitized}, here restricted to two-local correlations and optionally weighted by the two-qubit gate error rates of the selected QPU. Given the mutual-information matrix $J$ computed from the classical input features and the set of available hardware edges $\mathcal{E}$ induced by the QPU topology, the goal is to find a permutation $\pi$ assigning features to qubits so as to preserve as much pairwise statistical dependence as possible on physically available interactions. In the two-body setting used here, the fitness function is
\begin{equation}
    F(\pi)
    =
    \sum_{(i,j)\in\mathcal{E}}
    w_{ij}
    J_{\pi(i),\pi(j)},
    \label{eq:ga_assignment_cd_ising}
\end{equation}
where $w_{ij}=1$ by default, or may be chosen from calibration data to favor lower-error two-qubit links. The genetic algorithm evolves candidate permutations using tournament selection, order crossover, mutation, and elitism. In the multilayer CD-Ising-glass construction, this assignment is applied independently to each feature block, so that feature pairs with high mutual information are preferentially placed on neighboring qubits in each encoding layer.

\subsection{Heisenberg Map}

The Heisenberg map was the first approach to explicitly adopt the terminology of projected quantum feature maps~\cite{ciceri2025enhanced}. It is closely connected to the projected quantum kernel framework, since Huang et al.~\cite{huang2021power} considered the evolution of the one-dimensional Heisenberg Hamiltonian as one of their quantum data embeddings. The proposal in Ref.~\cite{ciceri2025enhanced} follows this physical motivation and uses the Heisenberg Hamiltonian as a data-dependent quantum feature map. The corresponding chain Hamiltonian can be written as
\begin{equation}
    H_{J}
    =
    \frac{1}{4}
    \sum_{j=1}^{n-1}
    J_j
    \left(
    X_{j}X_{j+1}
    +
    Y_{j}Y_{j+1}
    +
    Z_{j}Z_{j+1}
    \right),
    \label{eq:heisenberg_hamiltonian}
\end{equation}
where $n$ is the number of qubits, $X_j,Y_j,Z_j$ are Pauli operators acting on qubit $j$, and $J_j$ denotes the exchange coupling associated with the edge $(j,j+1)$. In this PQFM setting, these couplings are not the mutual-information couplings $J_{ij}$ used in the CD-Ising-glass map. This is one of the main differences between the two approaches. Here, the data are encoded in the interaction terms rather than in local fields, and no pairwise mutual-information encoding is used. So, to avoid confusion, following \cite{ciceri2025enhanced} we denote the Heisenberg encoding parameters by \(\eta_j\),
\begin{equation}
    \eta_j
    =
    2\pi \tanh\left(\frac{h_j}{3}\right),
    \qquad
    r=1,\ldots,p,
    \label{eq:heisenberg_tanh_scaling}
\end{equation}
where $h_j$ is the normalized value of the classical feature as defined in the last section. These values are then assigned to nearest-neighbor chain edges and used as the effective exchange couplings in the Heisenberg interaction terms. Here, we follow the same approach used in Ref.~\cite{ferenczi2025credit}, where multiple features per qubit are encoded by increasing the number of layers, or feature blocks, in the circuit. Thus, for an edge \((j,j+1)\) in block \(b\), the elementary interaction block is written as
\begin{equation}
    U_{j}^{(b)}(\mathbf{x})
    =
    \exp
    \left[
    -i\gamma\,
    \eta_j^{(b)}
    \left(
    X_{j}X_{j+1}
    +
    Y_{j}Y_{j+1}
    +
    Z_{j}Z_{j+1}
    \right)
    \right],
    \label{eq:heisenberg_edge_unitary}
\end{equation}
where \(\eta_j^{(b)}\) denotes the transformed classical feature assigned to edge \((j,j+1)\) in feature block \(b\), and \(\gamma\) is an effective scaling parameter that includes the global Heisenberg scale and the normalization by the number of feature blocks.

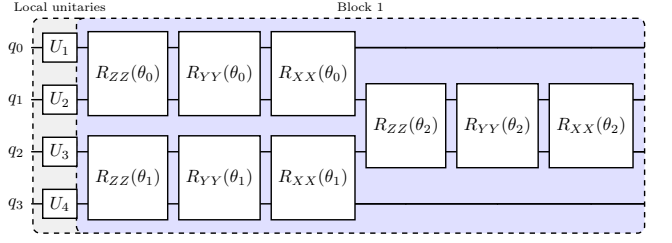
\begin{figure}[t]
\centering
\resizebox{\columnwidth}{!}{%
\begin{quantikz}[row sep=0.40cm, column sep=0.22cm]
\lstick{$q_0$} &
\gate{U_1}
\gategroup[wires=4,steps=1,style={dashed, rounded corners, fill=gray!12, inner xsep=2pt},background]
{{\scriptsize Local unitaries}} &
\gate[wires=2]{R_{ZZ}(\theta_{0})}
\gategroup[wires=4,steps=6,style={dashed, rounded corners, fill=blue!12, inner xsep=3pt},background]
{{\scriptsize Block 1}} &
\gate[wires=2]{R_{YY}(\theta_{0})} &
\gate[wires=2]{R_{XX}(\theta_{0})} &
\qw &
\qw &
\qw &
\qw \\
\lstick{$q_1$} &
\gate{U_2} &
&
&
&
\gate[wires=2]{R_{ZZ}(\theta_{2})} &
\gate[wires=2]{R_{YY}(\theta_{2})} &
\gate[wires=2]{R_{XX}(\theta_{2})} &
\qw \\
\lstick{$q_2$} &
\gate{U_3} &
\gate[wires=2]{R_{ZZ}(\theta_{1})} &
\gate[wires=2]{R_{YY}(\theta_{1})} &
\gate[wires=2]{R_{XX}(\theta_{1})} &
&
&
&
\qw \\
\lstick{$q_3$} &
\gate{U_4} &
&
&
&
\qw &
\qw &
\qw &
\qw
\end{quantikz}
}
\captionsetup{justification=raggedright,singlelinecheck=false}
\caption{\justifying
Example of the 4-qubit encoding 3 features using Heisenberg PQFM. Each qubit first undergoes a local Haar-random unitary transformation, denoted by $U_1,\dots,U_4$. The entangling stage, grouped here as Block~1, then applies the sequence of two-qubit rotations $R_{ZZ}$, $R_{YY}$, and $R_{XX}$ to the disjoint pairs $(q_0,q_1)$ and $(q_2,q_3)$, followed by the same sequence on the central pair $(q_1,q_2)$. More blocks can be added for encoding more features. The corresponding rotation angles are denoted by $\theta_0$, $\theta_1$, and $\theta_2$, where $\theta_i = 2\gamma\eta_i$, $\gamma$ is a scaling parameter, and $\eta_i$ is defined in Eq.~\ref{eq:heisenberg_tanh_scaling}.
}
\label{fig:heisenberg_brickwork_4q}
\end{figure}

The full feature map is implemented as an even--odd product of nearest-neighbor interaction blocks. Since each feature block contains \(n-1\) nearest-neighbor edges, it can encode \(n-1\) classical features. For the implementation of the unitary operator defined in Eq. \ref{eq:heisenberg_edge_unitary} a first-order Trotter product is used between the two-local terms. Feature vectors with \(p>n-1\) entries are therefore encoded by stacking
\begin{equation}
    B
    =
    \left\lceil
    \frac{p}{n-1}
    \right\rceil
\end{equation}
feature blocks. So, the corresponding full Heisenberg feature-map unitary can be written as
\begin{equation}
    U_H(\mathbf{x})
    =
    \left[
    \prod_{b=1}^{B}
    \left(
    \prod_{\substack{j=1 \\ j\,\mathrm{odd}}}^{n-1}
    U_{j}^{(b)}(\mathbf{x})
    \prod_{\substack{j=1 \\ j\,\mathrm{even}}}^{n-1}
    U_{j}^{(b)}(\mathbf{x})
    \right)
    \right]^{R}.
    \label{eq:heisenberg_feature_unitary}
\end{equation}
Here, \(R\) is the number of repetitions that, in general, are used as $1$ or $2$ for the called short and long circuits, respectively, as defined in \cite{ciceri2025enhanced}. The notation \(\eta_j^{(b)}\) makes explicit that the same physical edge can encode different classical features in different blocks. Features not present in the last incomplete block are padded with zero-valued parameters. The encoded quantum state, in turn, is then \(\ket{\psi_{\mathbf{x}}}=U_H(\mathbf{x})\ket{\psi_0}\), where the fiducial state \(\ket{\psi_0}\) is prepared by applying fixed Haar-random single-qubit unitaries to the initial computational state \(\ket{0}^{\otimes n}\). Figure \ref{fig:heisenberg_brickwork_4q} illustrates the idea for $B=1$.

Finally, quantum features are extracted from local Pauli expectation values,
\begin{equation}
    x'_{a,j}
    =
    \bra{\psi_{\mathbf{x}}}
    \sigma^{a}_{j}
    \ket{\psi_{\mathbf{x}}},
    \qquad
    a\in\{X,Y,Z\},
\end{equation}
with optional two-local same-axis observables,
\begin{equation}
    x'_{aa,jk}
    =
    \bra{\psi_{\mathbf{x}}}
    \sigma^{a}_{j}
    \sigma^{a}_{k}
    \ket{\psi_{\mathbf{x}}},
    \qquad
    a\in\{X,Y,Z\}.
\end{equation} 

It is worth mentioning that the one-dimensional Heisenberg structure makes this PQFM particularly amenable to matrix-product-state (MPS) simulation. Indeed, Refs.~\cite{ciceri2025enhanced, ferenczi2025credit} reports simulations with more than 100 qubits under a one-dimensional entanglement constraint. This choice contrasts with more hardware-native constructions that exploit a larger fraction of the target QPU connectivity and then are usually classically hard to simulate. Although classical hardness is not sufficient to guarantee improved predictive performance ~\cite{huang2021power}, it is more naturally aligned with strong claims of quantum advantage, since the evaluation of the corresponding quantum features may itself require genuinely quantum computational resources~\cite{Liu2021,Glick2024}. This motivates the generalized hardware-aware Heisenberg-type maps introduced in the next section. Instead of restricting the interaction graph to a one-dimensional chain, these constructions use the available two-qubit connectivity of the quantum processor to define richer projected feature maps, closer in spirit to the CD-Ising-glass PQFM.

\subsection{Generalized two-qubit Hamiltonian map}
The previous two subsections described the CD-Ising-glass and Heisenberg PQFMs, which are implemented in \texttt{pqfmlib} as reference Hamiltonian-based feature maps. The library also introduces the main methodological contribution of this work, a generalized two-qubit Hamiltonian PQFM framework called \texttt{XYZProjectiveQFM}. This framework combines key elements of the reference constructions while allowing more general combinations of two-qubit Pauli interactions. To motivate this construction, we start from a general two-qubit Hamiltonian expanded in the Pauli basis as defined in
\cite{Klassen2019twolocalqubit}
\begin{equation}
    H
    =
    \sum_{\mu,\nu\in\{I,X,Y,Z\}}
    a_{\mu\nu}\,
    \sigma^\mu \otimes \sigma^\nu,
    \qquad
    a_{\mu\nu}\in\mathbb{R},
\end{equation}
where we may take \(a_{II}=0\) without loss of generality. Inspired by the CD-Ising-glass structure (Eq. \ref{eq:ising_glass_hamiltonian}), in which normalized input features are encoded in one-local fields and pairwise mutual-information estimates are encoded in two-local interaction terms, we define the following family of two-qubit Hamiltonian-based PQFMs:
\begin{equation}
    H_{\mathrm{XYZ}}(\mathbf{x})
    =
    \sum_{i=1}^{n}
    \sum_{a\in\mathcal{A}}
    h_i^{(a)}
    \sigma_i^a
    +
    \sum_{(i,j)\in\mathcal{E}}
    \sum_{(a,b)\in\mathcal{P}}
    J_{ij}^{(ab)}
    \sigma_i^a\sigma_j^b ,
    \label{eq:xyz_general_hamiltonian}
\end{equation}
Here, \(\mathcal{A}\subseteq\{x,y,z\}\) denotes the set of active Pauli axes used for local feature encoding, and \(\mathcal{E}\) is the interaction graph on which two-body terms are implemented. In hardware executions, \(\mathcal{E}\) is chosen from the coupling topology of the selected QPU, possibly taking calibration information into account. The set \(\mathcal{P}\subseteq\mathcal{A}\times\mathcal{A}\) specifies the active two-body Pauli channels. In the present implementation, this is done by enabling one of three channel families:
\begin{equation}
\begin{aligned}
    \mathcal{P}
    &\in
    \left\{
    \mathcal{P}_{\mathrm{diag}},
    \mathcal{P}_{\mathrm{cross}},
    \mathcal{P}_{\mathrm{diag}}\cup\mathcal{P}_{\mathrm{cross}}
    \right\},
    \\
    \mathcal{P}_{\mathrm{diag}}
    &=
    \left\{
    (a,a): a\in\mathcal{A}
    \right\},
    \\
    \mathcal{P}_{\mathrm{cross}}
    &=
    \left\{
    (a,b): a,b\in\mathcal{A},\ a\neq b
    \right\}.
\end{aligned}
\label{eq:xyz_channel_families}
\end{equation}
Thus, diagonal-only, cross-axis-only, and mixed interaction structures are supported. This construction ties the available two-body channels to the axes used for local encoding. For example, if \(\mathcal{A}=\{x,y\}\), the diagonal-only configuration contains both \(XX\) and \(YY\) channels, while the cross-axis configuration contains \(XY\) and \(YX\). Consequently, a Hamiltonian with local \(X\) and \(Y\) fields but only an \(XX\) interaction channel is not part of the current \texttt{XYZProjectiveQFM} parameterization.

The main purpose of this family is to make multi-feature encoding possible through the Pauli-axis degree of freedom, something that was mentioned in \cite{ciceri2025enhanced} as a natural possibility in the Heisenberg map case. In this case, different axes of the same qubit can receive different classical features, so that a register with \(n\) qubits and \(|\mathcal{A}|\) active axes can encode up to \(n|\mathcal{A}|\) features per layer. This axis-based encoding is complementary to the multilayer strategy used in the reference CD-Ising-glass and Heisenberg maps, where additional features are accommodated by increasing the number of encoding layers. The \texttt{XYZProjectiveQFM} can also combine both mechanisms, using multiple axes within each layer and additional layers when the number of input features exceeds the per-layer capacity. Consequently, one could in principle encode into $n$ qubits, $|\mathcal{A}|$ active axis and $L$ encoding layers up to :
\begin{equation}
    C_{\mathrm{multi\text{-}axis}}
    =
    n\,|\mathcal{A}|\,L .
    \label{eq:xyz_multiaxis_capacity}
\end{equation}
For example, with \(n=4\) qubits, \(\mathcal{A}=\{x,y,z\}\), and \(L=2\) layers, the multi-axis mode can encode up to $C_{\mathrm{multi\text{-}axis}}=4\times 3\times 2=24$ distinct classical features. In this case, each qubit carries three features per layer, one encoded along each Pauli axis, so each layer accommodates twelve features.

\begin{figure*}[!t]
\centering
\includegraphics[width=\textwidth]{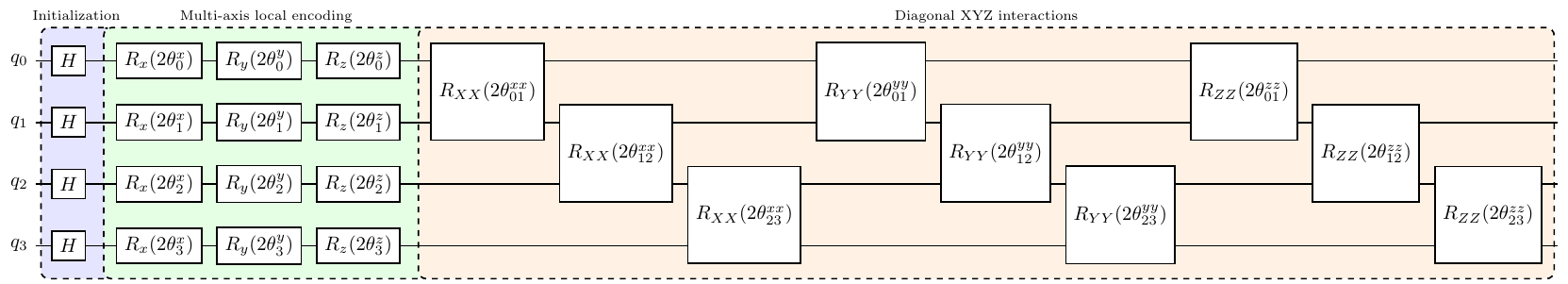}
\caption{\justifying
Circuit representation of a of a single-layer diagonal XYZ PQFM with twelve features mapped onto a four-qubit register by a multi-axis feature encoding. After Hadamard initialization, each qubit receives three local rotations, \(R_x\), \(R_y\), and \(R_z\), allowing three distinct normalized classical features to be encoded per qubit through the active Pauli axes. The subsequent two-qubit rotations implement the diagonal interaction channels \(XX\), \(YY\), and \(ZZ\), with angles determined by mutual-information coefficients between the features assigned to the corresponding qubit-axis slots. The nearest-neighbor chain shown here is illustrative; in the hardware-aware implementation, the interaction pattern is determined by the processor coupling map. 
}
\label{fig:circuit-4q-1layers_xyz}
\end{figure*}

This axis-resolved encoding also requires a corresponding extension of the hardware-aware feature-assignment procedure. In the one-feature-per-qubit case, the assignment reduces to the qubit-level permutation described in Eq.~\eqref{eq:ga_assignment_cd_ising}. For multi-axis encoding, however, features must be assigned to qubit-axis slots \((i,a)\), where \(i\) labels the qubit and \(a\in\mathcal{A}\) labels the Pauli axis used for local encoding. The available two-qubit hardware edges are therefore lifted to slot pairs \(((i,a),(j,b))\), with \((i,j)\in\mathcal{E}\) and \((a,b)\in\mathcal{P}\). The corresponding fitness function can be written as
\begin{equation}
    F(\pi)
    =
    \sum_{(i,j)\in\mathcal{E}}
    \sum_{(a,b)\in\mathcal{P}}
    w_{ij}\,
    J_{\pi(i,a),\pi(j,b)},
    \label{eq:ga_assignment_multiaxis}
\end{equation}
where \(\pi(i,a)\) denotes the classical feature assigned to the slot \((i,a)\). This preserves the same permutation-based genetic core used in the CD-Ising-glass assignment, while enlarging the search space from qubits to axis-resolved encoding slots. As a result, pairwise statistical dependencies between features assigned to different Pauli axes can also be matched to hardware-compatible two-qubit interactions, provided that those features are assigned to axes on different qubits. Compared with a purely layer-based multi-feature strategy, the multi-axis assignment can allow to encode more pairwise mutual information, being probably a more interesting strategy to multi-feature encoding PQFM in this framework as we will see empirically in the Section \ref{sec:numerical_results}. 

By contrast, one could also choose to encode the same feature across all active axes of a given qubit, intending to explore richer representations of the same information. In that case number of distinct encoded features is
\begin{equation}
    C_{\mathrm{shared}}
    =
    nL ,
    \label{eq:xyz_shared_capacity}
\end{equation}
while \(|\mathcal{A}|\) controls how many Pauli directions are used to encode each feature. Both choices are available in \texttt{pqfmlib}.

All the concrete maps used in this work are recovered as special cases of Eq.~\eqref{eq:xyz_general_hamiltonian}. In particular, choosing \(\mathcal{A}=\{x,y,z\}\) and \(\mathcal{P}=\{(x,x),(y,y),(z,z)\}\) gives a diagonal XYZ map,
\begin{equation}
\begin{aligned}
    H_{\mathrm{diag}}
    &=
    \sum_i
    \left(
    h_i^{(x)}X_i
    +
    h_i^{(y)}Y_i
    +
    h_i^{(z)}Z_i
    \right)
    \\
    &\quad
    +
    \sum_{(i,j)\in\mathcal{E}}
    \left(
    J_{ij}^{(xx)}X_iX_j
    +
    J_{ij}^{(yy)}Y_iY_j
    +
    J_{ij}^{(zz)}Z_iZ_j
    \right),
\end{aligned}
\label{eq:xyz_diag_hamiltonian}
\end{equation}
which is structurally related to anisotropic Heisenberg or XYZ spin Hamiltonians~\cite{langari2004quantum,claeys2019integrable} or with the isotropic Heisenberg if one choose the encode the same feature in all axes ($ h_i^{(x)} =  h_i^{(y)} =  h_i^{(z)}$ and $J_{ij}^{(xx)} = J_{ij}^{(yy)} = J_{ij}^{(zz)}$) \cite{burkard2025recipes}. A simple example of the circuit implementation of this Hamiltonian can be seen in Fig. \ref{fig:circuit-4q-1layers_xyz}. Choosing instead only off-diagonal Pauli channels, \(\mathcal{P}=\{(a,b):a,b\in\mathcal{A},a\neq b\}\), gives a cross-axis XYZ map. In particular, for $\mathcal{P}=\{(x,y), (y,x), (x,z), (z,x), (y,z), (z,y)\}$ we have
\begin{equation}
\begin{aligned}
    H_{\mathrm{cross}}
    &=
    \sum_i
    \left(
    h_i^{(x)}X_i
    +
    h_i^{(y)}Y_i
    +
    h_i^{(z)}Z_i
    \right)
    \\
    &\quad
    +
    \sum_{(i,j)\in\mathcal{E}}
    \Big(
    J_{ij}^{(xy)}X_iY_j
    +
    J_{ij}^{(yx)}Y_iX_j
    +
    J_{ij}^{(xz)}X_iZ_j
    \\
    &\quad
    +
    J_{ij}^{(zx)}Z_iX_j
    +
    J_{ij}^{(yz)}Y_iZ_j
    +
    J_{ij}^{(zy)}Z_iY_j
    \Big).
\end{aligned}
\label{eq:xyz_cross_hamiltonian}
\end{equation}
This construction is inspired by the CD-Ising-glass map and can be viewed as a generalized Hamiltonian structure that retains its distinction between local feature fields and correlation-informed two-body couplings. In particular, choosing \(\mathcal{A}=\{y,z\}\) and \(\mathcal{P}=\{(y,z),(z,y)\}\) yields a cross-axis interaction structure closely related to the counterdiabatic two-qubit blocks used as the PQFM in the CD-Ising-glass construction (Eq. \ref{eq:cd_ising_final_hamiltonian}). Finally, taking \(\mathcal{P}=\mathcal{A}\times\mathcal{A}\) gives the full two-local XYZ map, containing both diagonal and cross-axis Pauli interactions. This has the same operator structure as general bilinear spin Hamiltonians with exchange tensors, as used in digital quantum simulation recipes for lattice spin systems~\cite{burkard2025recipes} and in general treatments of spin Hamiltonians in magnetic materials~\cite{li2021spin}. In this case, we will have 
\begin{equation}
    J_{ij} =
    \begin{pmatrix}
    J_{ij}^{(xx)} & J_{ij}^{(xy)} & J_{ij}^{(xz)} \\
    J_{ij}^{(yx)} & J_{ij}^{(yy)} & J_{ij}^{(yz)} \\
    J_{ij}^{(zx)} & J_{ij}^{(zy)} & J_{ij}^{(zz)}
    \end{pmatrix},
\label{eq:xyz_full_jij}   
\end{equation}
coupling terms to encode the pairwise mutual information and 
\begin{equation}
    \mathbf{h}_i = \left(h_i^{(x)},\, h_i^{(y)},\, h_i^{(z)}\right),
    \label{eq:xyz_full_hi}
\end{equation}
local terms to encode the normalized feature. It is worth mentioning that in the circuit implementation of all these Hamiltonians available by the two-qubit family follows the same path as the other PQFMs, a first-order Trotter approximation with only one step, although more steps can be used.

\section{Methodology}
\label{sec:lib_cross_validation}
This section describes the experimental protocol used to benchmark the PQFMs considered in this work. We present the datasets, preprocessing steps, quantum-feature generation settings, classical learning pipeline, and the paired statistical analysis used to compare quantum and classical feature representations.

\subsection{Datasets and preprocessing}

To evaluate the performance of the different PQFMs, we selected four medical and biomedical binary-classification datasets. These datasets were chosen to cover different regimes of sample size, feature dimension, and application domain, while remaining compatible with the quantum-feature extraction strategies considered in this work.

\begin{itemize}
\item \textbf{Molecular Toxicity - CRY1}: a dataset composed of $171$ small molecules designed to target functional domains of the core clock protein CRY1 \cite{gul2021structure}. The original dataset contains 1203 molecular descriptors. To make the problem compatible with the quantum-hardware experiments considered here, we selected the top 156 features according to their mutual information with the target variable.

\item \textbf{Breast Cancer Radiology - I-SPY1}: a breast-cancer radiomics dataset obtained from \texttt{radMLBench}, a curated benchmark collection of tabular radiomic datasets  \cite{demircioglu2024radmlbench}. The dataset is derived from the TCIA I-SPY1 collection, which contains multi-center breast DCE-MRI data and segmentations from patients in the I-SPY 1/ACRIN 6657 trials \cite{newitt2016ispy1}. The tabular version used here contains $161$ patient samples and $370$ radiomic features, with a binary target corresponding to recurrence-free survival status as provided in the \texttt{radMLBench} tabular benchmark.

\item \textbf{Breast Cancer Pathology - Wisconsin Diagnostic}: a classical diagnostic breast-cancer dataset based on digitized fine-needle aspirate images of breast masses  \cite{street1993nuclear}. It contains $569$ samples described by $30$ real-valued nuclear morphology features, with labels corresponding to benign or malignant tumors.

\item \textbf{Cardiac Disease Clinical - Cleveland}: a clinical dataset derived from patients evaluated for coronary artery disease at the Cleveland Clinic \cite{detrano1989international}. The version used here contains $303$ samples and $13$ multidimensional features, including clinical, laboratory and results of medical imaging data with a binarized target indicating absence or presence of heart disease based on angiogram results.
\end{itemize}
\FloatBarrier

\noindent In all experiments, input features were standardized to zero mean and unit variance, and missing values were imputed using the median when applicable. The first two datasets, Molecular Toxicity and Breast Cancer Radiology, were used for real quantum-hardware experiments. The Breast Cancer Pathology and Cardiac Disease datasets were used in \texttt{statevector} simulation experiments, allowing us to explore additional hybrid classical-quantum feature-selection strategies in a controlled setting. For clarity, Table~\ref{tabdataset} summarizes the main characteristics and role of each dataset in the benchmark.

\begin{table*}[t]
\caption{Summary of the datasets and experimental regimes considered in this work.}
\centering
\small
\renewcommand{\arraystretch}{1.20}
\setlength{\tabcolsep}{5pt}

\resizebox{\textwidth}{!}{
\begin{tabular}{@{}lcccl@{}}
\hline
\rowcolor{gray!15}[0.8\tabcolsep][0.8\tabcolsep]
\textbf{Dataset} &
\textbf{Samples} &
\textbf{Original features} &
\textbf{Used features} &
\textbf{Experimental regime} \\
\hline
\rule{0pt}{3.0ex}
Molecular Toxicity
& 171 & 1203 & 156
& IBM QPU \\
\rule{0pt}{3.0ex}
Breast Cancer Radiology
& 161 & 370 & 370
& IBM QPU \\
\rule{0pt}{5.0ex}
\shortstack[l]{Breast Cancer Pathology}
& \raisebox{0.4\normalbaselineskip}{569}
& \raisebox{0.4\normalbaselineskip}{30}
& \raisebox{0.4\normalbaselineskip}{30}
& \shortstack[l]{Statevector simulation\\SHAP-based feature selection} \\
\rule{0pt}{5.0ex}
\shortstack[l]{Cardiac Disease - Cleveland}
& \raisebox{0.4\normalbaselineskip}{303}
& \raisebox{0.4\normalbaselineskip}{13}
& \raisebox{0.4\normalbaselineskip}{13}
& \shortstack[l]{Statevector simulation\\PCA-based dimensionality reduction} \\
\hline
\end{tabular}\label{tabdataset}
}
\end{table*}

\subsection{Quantum feature generation and hardware settings}

For quantum-feature generation, we used the CD-Ising-glass and Heisenberg maps as reference Hamiltonian-based PQFMs and compared them with the generalized two-qubit Hamiltonian family, including diagonal, cross-axis, and full two-local Pauli interaction structures. Within the multi-feature encoding setting, we compared two strategies: increasing the number of encoding blocks to accommodate more features, and encoding multiple features across different Pauli axes within the same block. We also evaluated \textit{shared-feature} configurations, in which the same classical feature is encoded along different Pauli axes.

For the Molecular Toxicity dataset, we compared three IBM quantum processors, \texttt{ibm\_marrakesh}, \texttt{ibm\_fez}, and \texttt{ibm\_kingston}, accessed through IBM's open-access plan. All hardware executions used $4096$ shots, with no quantum error correction or error-mitigation techniques. For the Breast Cancer Radiology dataset, we used only \texttt{ibm\_kingston}. This choice was made based on the Molecular Toxicity hardware benchmark and to reduce QPU resource usage, since \texttt{ibm\_kingston} provided the best overall performance among the tested processors.

For the Breast Cancer Pathology and Cardiac Disease datasets, quantum features were generated using the \texttt{statevector} mode of Qiskit's \texttt{AerSimulator}. Nevertheless, these simulations were constructed using the same hardware-aware feature-assignment procedure adopted for real-device executions. This choice provides a natural way to select high-mutual-information feature pairs under hardware-connectivity constraints and makes the simulated experiments more directly comparable to the hardware setting.

For each PQFM configuration, the output of the quantum-feature extraction stage consists of classical feature matrices formed from measured or simulated expectation values. Depending on the map, we considered one-local observables, such as single-qubit Pauli expectation values, and two-local observables, such as pairwise Pauli correlations. For the Heisenberg map, local features were extracted along the $X$, $Y$, and $Z$ axes. For the CD-Ising-glass map, we focused on $Z$-axis observables. For the generalized two-qubit Hamiltonian family, the available observables depend on the active Pauli axes and interaction channels of each configuration.

\subsection{Classical learning pipeline}

For the classical learning pipeline, we used a nested cross-validation protocol designed to separate model selection from external performance estimation. The outer loop consisted of repeated stratified 5-fold cross-validation with 10 repetitions, yielding 50 independent outer evaluations. In each outer fold, the outer test set was kept completely held out, while the corresponding outer training set was used for model selection. Hyperparameters were optimized through an inner stratified 80/20 holdout split implemented with \texttt{PredefinedSplit} inside \texttt{GridSearchCV}, using AUC as the selection criterion. After the best hyperparameter configuration was selected, the pipeline was refitted on the full outer training set and then evaluated once on the untouched outer test set. Final performance was summarized over the 50 outer folds, ensuring that hyperparameter selection and performance estimation remained separated.

\begin{table}[H]
\centering
\small
\setlength{\tabcolsep}{6pt}
\renewcommand{\arraystretch}{1.15}
\begin{tabular}{ll}
\toprule
\textbf{Hyperparameter} & \textbf{Search space} \\
\midrule
\texttt{n\_estimators} & $\{80,100,120,140,160\}$ \\
\texttt{max\_depth} & $\{2,5,7,9,11\}$ \\
\texttt{criterion} & $\{\texttt{friedman\_mse},\texttt{squared\_error}\}$ \\
\bottomrule
\end{tabular}
\caption{\justifying Hyperparameter search space used for the Gradient Boosting classifier.}
\label{tab:gb_hyperparameters}
\end{table}

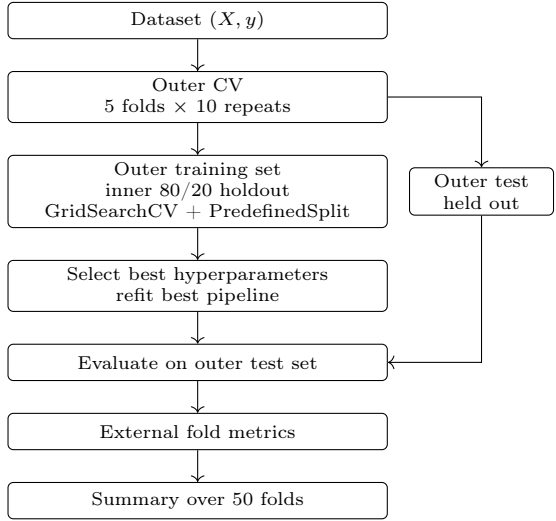
\begin{figure}[t]
\centering
\begin{tikzpicture}[
    node distance=0.42cm,
    box/.style={
        rectangle,
        draw,
        rounded corners=2pt,
        align=center,
        font=\scriptsize,
        text width=0.56\columnwidth,
        minimum height=0.48cm,
        inner sep=2.5pt
    },
    sidebox/.style={
        rectangle,
        draw,
        rounded corners=2pt,
        align=center,
        font=\scriptsize,
        text width=0.20\columnwidth,
        minimum height=0.48cm,
        inner sep=2.5pt
    },
    arrow/.style={->, line width=0.4pt}
]

\node[box] (data) {Dataset $(X,y)$};

\node[box, below=of data] (outercv)
{Outer CV\\5 folds $\times$ 10 repeats};

\node[box, below=of outercv] (inner)
{Outer training set\\
inner 80/20 holdout\\
GridSearchCV + PredefinedSplit};

\node[box, below=of inner] (refit)
{Select best hyperparameters\\
refit best pipeline};

\node[box, below=of refit] (eval)
{Evaluate on outer test set};

\node[box, below=of eval] (metrics)
{External fold metrics};

\node[box, below=of metrics] (summary)
{Summary over 50 folds};

\node[sidebox, right=0.28cm of inner] (test)
{Outer test\\held out};

\draw[arrow] (data) -- (outercv);
\draw[arrow] (outercv) -- (inner);
\draw[arrow] (inner) -- (refit);
\draw[arrow] (refit) -- (eval);
\draw[arrow] (eval) -- (metrics);
\draw[arrow] (metrics) -- (summary);

\draw[arrow] (outercv.east) -| (test.north);
\draw[arrow] (test.south) |- (eval.east);

\end{tikzpicture}
\caption{\justifying Nested validation protocol used for model selection and external performance estimation.}
\label{fig:nested_cv_pipeline}
\end{figure}

We focused on a Gradient Boosting classifier with random seed fixed to $42$. The hyperparameter search space is reported in Table~\ref{tab:gb_hyperparameters}. AUC was used as the primary metric for model selection, while F1 score, precision, recall, and accuracy were used as complementary diagnostic metrics. The same nested validation protocol was applied to classical, quantum-only, and hybrid classical--quantum feature representations. A schematic representation of the validation procedure is shown in Fig.~\ref{fig:nested_cv_pipeline}.

\subsection{Hybrid feature-selection strategies in simulation}

While in the quantum-hardware experiments we focused on evaluating whether quantum-generated features, including both one-local and two-local observables, could enhance model performance by themselves, in the simulation experiments we also explored hybrid classical--quantum feature spaces. For the Breast Cancer Pathology dataset, we used SHAP-based feature selection \cite{lundberg2017shap,lundberg2020tree} to identify the most relevant quantum features and the most relevant features in the combined classical--quantum space. Based on our preliminary experiments, we retained the top $30$ SHAP-ranked features, since this setting provided the best overall performance, although comparable results were also observed when using a larger number of features. For the Cardiac Disease dataset, we used principal component analysis (PCA) as an alternative dimensionality-reduction strategy \cite{jolliffe2002pca,jolliffe2016pca}, retaining $13$ principal components according to the same empirical criterion.

These two complementary approaches allow us to assess the usefulness of the quantum features from different perspectives. SHAP provides a supervised, model-based ranking of feature relevance, whereas PCA probes whether the enlarged feature space contains informative directions of variation. Importantly, the same nested cross-validation pipeline described above was preserved; only the feature-selection or dimensionality-reduction strategy was changed. This makes the comparison more robust, since improvements obtained under different selection mechanisms suggest that the quantum features can provide useful additional information for improving classical machine-learning performance.

\begin{figure*}[t]
  \centering
  \includegraphics[width=\textwidth]{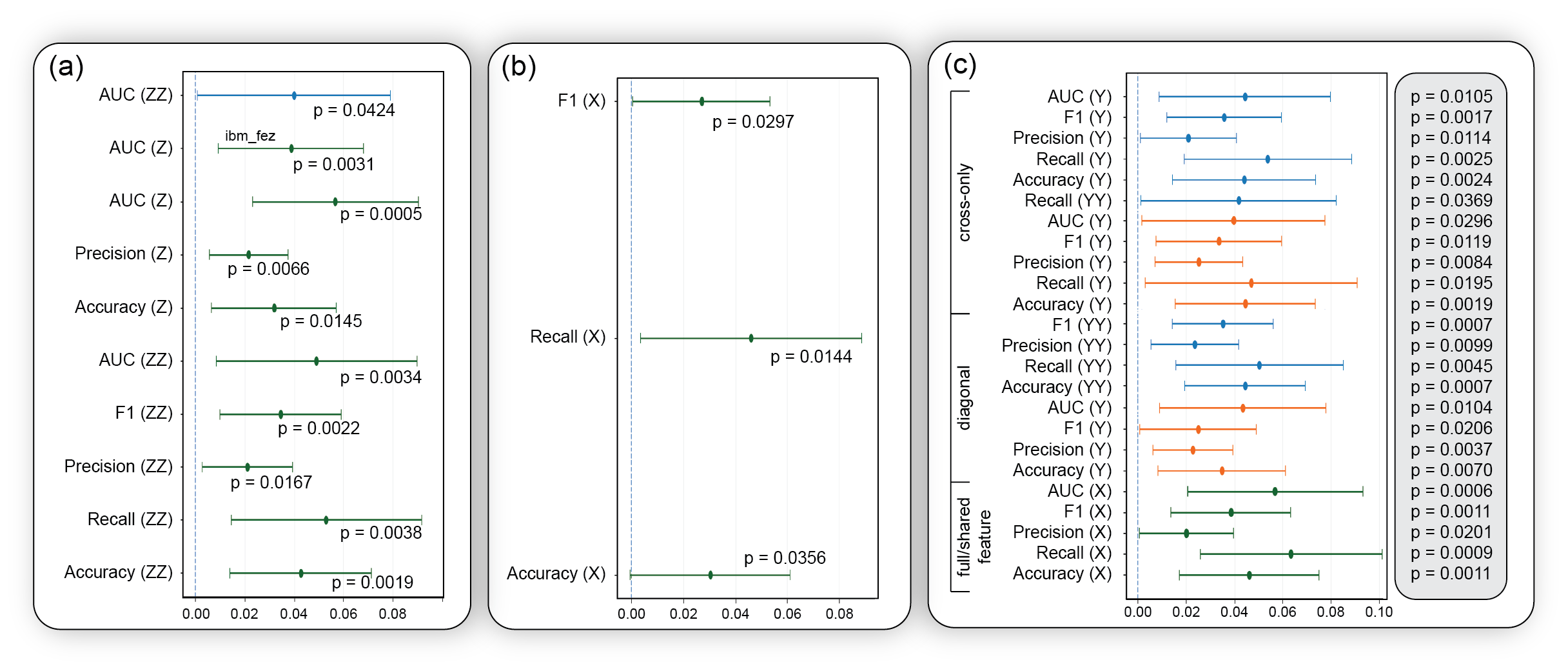}
    \caption{\justifying\label{fig:stat_tox}
    Paired statistical improvements over the classical baseline for the Molecular Toxicity dataset. Panel (a) shows the CD-Ising-glass PQFM, panel (b) the Heisenberg map, and panel (c) the XYZ feature-map family introduced in this work. The plotted intervals correspond to the mean paired deltas and 95\% confidence intervals. Colors indicate the qubit budget used in each configuration, with blue, orange, and green corresponding to 52, 78, and 156 qubits, respectively. The labels in parentheses indicate the measured observable used to extract the corresponding quantum features. Unless otherwise annotated, results correspond to \texttt{ibm\_kingston}; panel (b) uses \texttt{ibm\_marrakesh}.}
\end{figure*}

\subsection{Statistical analysis}
\label{sec:statistical_analysis}
To assess the statistical significance of our results, we performed, for all experiments, a paired fold-by-fold comparison against the corresponding classical baseline. For each quantum or hybrid classical--quantum configuration, and for each metric, we computed the paired difference on every outer validation split as
\begin{equation}
\Delta_i = M_i^{\mathrm{case}} - M_i^{\mathrm{classical}},
\end{equation}
where $M_i^{\mathrm{case}}$ is the metric obtained by the evaluated configuration on outer fold $i$, and $M_i^{\mathrm{classical}}$ is the corresponding classical-baseline metric on the same fold. This pairing ensures that each comparison uses the same train-test split, reducing the effect of fold-to-fold variability.

We summarized the distribution of paired deltas over the 50 outer folds by its mean and computed a 95\% confidence interval for the mean delta using the Student-$t$ distribution. In addition, we applied a one-sided Wilcoxon signed-rank test with alternative ``greater'' to test whether the evaluated configuration tended to outperform the classical baseline across folds \cite{wilcoxon_1,GibbonsChakraborti2010}. The same procedure was used for the quantum-only experiments and for the simulation experiments involving SHAP and PCA. Detailed paired-delta statistics and Wilcoxon results are reported in Appendix~\ref{app:paired_statistical_analysis}.

When interpreting the results, we use AUC as the primary model-selection metric and treat F1 score, precision, recall, and accuracy as complementary metrics that help characterize the behavior of each model. Improvements are considered statistically supported, or \textit{Consistent}, when the mean paired delta is positive, the 95\% confidence interval lies entirely above zero, and the one-sided Wilcoxon signed-rank test gives $p<0.05$. Cases with positive mean delta and Wilcoxon $p<0.05$, but with a confidence interval crossing zero, are reported as \textit{Promising}. Also, we consider as \textit{Suggestive} those cases with positive mean delta, confidence interval crossing zero, and $0.05 \leq p < 0.10$. Positive deltas with $p \geq 0.10$ and confidence intervals crossing zero are reported only as \textit{Positive trends}, without being interpreted as statistically supported improvements.

\section{Numerical Results and Analysis}
\label{sec:numerical_results}

We organize the results around three questions. First, we compare the reference CD-Ising-glass and Heisenberg PQFMs with the proposed generalized Hamiltonian maps on real quantum hardware. Second, we analyze whether multi-axis feature encoding provides a better strategy than increasing the number of encoding layers. Third, we test whether projected quantum features remain useful in higher-dimensional and hybrid classical--quantum settings.

To keep the main text focused on the main result and physical interpretation, some tables and part of the detailed paired statistical analyses are reported in the appendices.

\subsection{Real-hardware benchmark on the Molecular Toxicity dataset}

Our first study focused on analyzing the multi-feature encoding strategy for the Molecular Toxicity dataset using our version of CD-Ising-glass and three QPUs for $52$, $78$ and $156$ qubits, corresponding to $3$, $2$, and $1$ classical features per qubit, respectively. The resulting performance summaries are reported in Table~\ref{tab:cdising_toxicity_means_only} from Appendix \ref{app:toxicity_resuts}. 

The results show that the benefit of the CD-Ising-glass quantum features is not uniform across all hardware backends and encoding regimes, but becomes clearer as the number of available qubits increases. In the most constrained regimes, with $52$ and $78$ qubits, the improvements depend on the metric and backend, with only a limited number of cases showing statistically supported gains over the classical baseline (see Fig.~\ref{fig:stat_tox}(a)). This indicates that, although the CD-Ising-glass map is a useful reference for Hamiltonian-based quantum feature extraction, its statistically supported gains are relatively sparse. In the 52-qubit regime, the only statistically supported gain is observed for \texttt{ibm\_kingston}, where the two-body $\langle ZZ \rangle$ observable improves AUC over the classical baseline. The strongest and most consistent gains are observed in the one-feature-per-qubit regime with 156 qubits, especially for \texttt{ibm\_kingston}. In this setting, the one-body $\langle Z \rangle$ features improve AUC, precision, and accuracy, while the two-body $\langle ZZ \rangle$ features improve all reported metrics under the paired statistical criteria. These results indicate that CD-Ising-glass quantum features can provide useful complementary information for the classical classifier. Furthermore, in the compressed regimes, where we encode multiple features per qubit, the results suggest that the feature map is still promising, albeit with more modest improvements.

We performed the same analysis for the Heisenberg map on the Molecular Toxicity dataset. In this case, the extracted quantum features are restricted to one-local observables measured along the three Pauli axes, $X$, $Y$, and $Z$. In Table~\ref{tab:heisenberg_toxicity_best_xyz}, we report, for each hardware configuration, only the axis that achieved the best AUC, with all metrics in the row corresponding to that same selected axis. Overall, the Heisenberg-map results are weaker than those obtained with the CD-Ising-glass map. This is also reflected in the paired statistical analysis based on confidence intervals for the fold-wise deltas and the Wilcoxon signed-rank test, where only a small number of metrics and configurations show statistically supported gains (see Fig.~\ref{fig:stat_tox}(b)). It is also worth noting that, because this PQFM requires measurements along all three Pauli axes, its execution requires approximately three times more QPU measurement time than single-axis feature maps.

We next consider the generalized two-qubit Hamiltonian in its diagonal (Eq.~\ref{eq:xyz_diag_hamiltonian}), cross-only (Eq.~\ref{eq:xyz_cross_hamiltonian}), and full configuration (Eqs.~\ref{eq:xyz_full_jij} and~\ref{eq:xyz_full_hi}). In this case, the generalized Hamiltonian maps show a more consistent pattern of improvements, especially considering the statistical tests in Fig. \ref{fig:stat_tox}(c). As shown in Table~\ref{tab:toxicity_xyz_kingston_cross_diag_shared}, the proposed $XYZ$ feature-map family improves over the classical baseline in most metrics and across all tested qubit budgets on \texttt{ibm\_kingston}. In the compressed regimes with $52$ and $78$ qubits, the multi-axis strategy provides clear gains, particularly through the one-body features in the cross-only setting, which improve all reported metrics relative to the classical baseline. Nevertheless, the diagonal setting also remains competitive, with its two-body features giving strong F1, precision, recall, and accuracy in the $52$-qubit case. Comparing with the results of the last two PQFMs that used the strategy of increasing the number of encoding blocks, for the Molecular Toxicity benchmark, the multi-axis strategy appears more effective than increasing only the number of encoding blocks.

The best overall performance is obtained in the 156-qubit full/shared-feature configuration, where the same feature is encoded along the three Pauli axes of each qubit. In this setting, the one-body observables reach an AUC of 0.6222, comparable to the best CD-Ising-glass result, while providing larger gains in F1 and recall, with values of 0.7520 and 0.8096, respectively.

\begin{table*}[t]
\centering
\scriptsize
\setlength{\tabcolsep}{2.8pt}
\renewcommand{\arraystretch}{1.15}
\resizebox{\textwidth}{!}{%
\begin{tabular}{lllcccccccccccccccc}
\toprule
\textbf{Algorithm} & \textbf{Qubits} & \textbf{Features} &
\multicolumn{5}{c}{\textbf{Classical}} &
\multicolumn{5}{c}{\textbf{SHAP Quantum}} &
\multicolumn{5}{c}{\textbf{SHAP Classical+Quantum}} &
\textbf{\%Q} \\
\\
\cmidrule(lr){4-8} \cmidrule(lr){9-13} \cmidrule(lr){14-18}
 &  &  &
\textbf{AUC} & \textbf{F1} & \textbf{Prec.} & \textbf{Rec.} & \textbf{Acc.} &
\textbf{AUC} & \textbf{F1} & \textbf{Prec.} & \textbf{Rec.} & \textbf{Acc.} &
\textbf{AUC} & \textbf{F1} & \textbf{Prec.} & \textbf{Rec.} & \textbf{Acc.} &
\\
\midrule

CD-Ising-glass & 20 & 30 &
0.9847 & 0.9643 & 0.9574 & 0.9717 & 0.9548 &
0.9825 & 0.9534 & 0.9522 & 0.9551 & 0.9415 &
\textbf{0.9880} & \textbf{0.9649} & \textbf{0.9596} & 0.9708 & \textbf{0.9557} &
35.93 \\

Heisenberg & 20 & 30 &
0.9847 & 0.9643 & 0.9574 & 0.9717 & 0.9548 &
\textbf{0.9911} & \textbf{0.9684} & \textbf{0.9637} & \textbf{0.9737} & \textbf{0.9601} &
0.9900 & 0.9672 & 0.9621 & 0.9728 & 0.9585 &
65.40 \\

XYZ-full/shared-feature & 20 & 30 &
0.9847 & 0.9643 & 0.9574 & 0.9717 & 0.9548 &
0.9596 & 0.9159 & 0.9055 & 0.9272 & 0.8933 &
\textbf{0.9885} & \textbf{0.9659} & \textbf{0.9581} & \textbf{0.9745} & \textbf{0.9567} &
43.00 \\

XYZ-diagonal/shared-feature & 20 & 30 &
0.9847 & 0.9643 & 0.9574 & 0.9717 & 0.9548 &
0.9653 & 0.9299 & 0.9200 & 0.9409 & 0.9109 &
\textbf{0.9884} & \textbf{0.9648} & \textbf{0.9603} & 0.9700 & \textbf{0.9555} &
47.47 \\

XY-cross-only & 15 & 30 &
0.9847 & 0.9643 & 0.9574 & 0.9717 & 0.9548 &
0.9584 & 0.9174 & 0.9019 & 0.9345 & 0.8944 &
\textbf{0.9896} & 0.9637 & 0.9551 & \textbf{0.9731} & 0.9539 &
32.27 \\

\bottomrule
\end{tabular}%
}
\caption{\justifying Breast Cancer Pathology results using SHAP-based feature selection with the top $30$ selected features and the nested validation protocol shown in Fig.~\ref{fig:nested_cv_pipeline}. We compare CD-Ising-glass, Heisenberg, XYZ-full/shared-feature, XYZ-diagonal/shared-feature, and XY-cross-only quantum feature-map variants. The classical baseline uses the original $30$ classical features. The ``SHAP Quantum'' block reports the performance obtained after selecting the top $30$ quantum features by SHAP. The ``SHAP Classical+Quantum'' block reports the performance obtained after selecting the top $30$ features from the union of classical and quantum features. The final column reports the average percentage of selected quantum features in this mixed classical+quantum SHAP set. All SHAP selections were performed without leakage within the nested validation protocol. Bold entries indicate, for each row and metric, the best quantum result whenever it exceeds the corresponding classical baseline.}
\label{tab:breast_cancer_shap_quantum_mixed}
\end{table*}

\begin{table*}[t]
\centering
\scriptsize
\setlength{\tabcolsep}{2.8pt}
\renewcommand{\arraystretch}{1.15}
\resizebox{\textwidth}{!}{%
\begin{tabular}{lllccccccccccccccc}
\toprule
\textbf{Algorithm} & \textbf{Qubits} & \textbf{Features} &
\multicolumn{5}{c}{\textbf{Classical}} &
\multicolumn{5}{c}{\textbf{PCA Quantum}} &
\multicolumn{5}{c}{\textbf{Classical+PCA Quantum}} \\
\cmidrule(lr){4-8} \cmidrule(lr){9-13} \cmidrule(lr){14-18}
 &  &  &
\textbf{AUC} & \textbf{F1} & \textbf{Prec.} & \textbf{Rec.} & \textbf{Acc.} &
\textbf{AUC} & \textbf{F1} & \textbf{Prec.} & \textbf{Rec.} & \textbf{Acc.} &
\textbf{AUC} & \textbf{F1} & \textbf{Prec.} & \textbf{Rec.} & \textbf{Acc.}
\\
\midrule

CD-Ising-glass & 13 & 13 &
0.8784 & 0.7797 & 0.8033 & 0.7626 & 0.8034 &
0.8654 & 0.7708 & 0.8014 & 0.7474 & 0.7957 &
0.8754 & \textbf{0.7894} & \textbf{0.8305} & 0.7554 & \textbf{0.8152} \\

Heisenberg & 14 & 13 &
0.8784 & 0.7797 & 0.8033 & 0.7626 & 0.8034 &
0.8790 & 0.7690 & 0.7855 & 0.7592 & 0.7911 &
\textbf{0.8855} & \textbf{0.7859} & 0.7992 & \textbf{0.7784} & \textbf{0.8056} \\

XYZ-cross/shared-feature & 13 & 13 &
0.8784 & 0.7797 & 0.8033 & 0.7626 & 0.8034 &
0.8387 & 0.7623 & 0.7552 & 0.7771 & 0.7793 &
\textbf{0.8873} & \textbf{0.7933} & 0.8026 & \textbf{0.7900} & \textbf{0.8109} \\

XYZ-cross-only/shared-feature & 13 & 13 &
0.8784 & 0.7797 & 0.8033 & 0.7626 & 0.8034 &
0.8304 & 0.7368 & 0.7452 & 0.7361 & 0.7601 &
\textbf{0.8821} & 0.7791 & 0.7880 & \textbf{0.7756} & 0.7984 \\

XYZ-diagonal/shared-feature & 13 & 13 &
0.8784 & 0.7797 & 0.8033 & 0.7626 & 0.8034 &
0.8189 & 0.7216 & 0.7356 & 0.7158 & 0.7482 &
0.8725 & 0.7690 & 0.7873 & 0.7582 & 0.7917 \\

\bottomrule
\end{tabular}%
}
\caption{\justifying Cardiac Disease results using PCA-based dimensionality reduction on quantum features with $13$ principal components and the nested validation protocol shown in Fig.~\ref{fig:nested_cv_pipeline}. The outer loop used stratified $5$-fold cross-validation with $10$ repetitions, yielding $50$ external test evaluations, while model selection was performed within each outer training set through an inner stratified $80/20$ holdout split using \texttt{PredefinedSplit} and \texttt{GridSearchCV}. We compare CD-Ising-glass, Heisenberg, XYZ-cross/shared-feature, XYZ-cross-only/shared-feature, and XYZ-diagonal/shared-feature quantum feature-map variants. The classical baseline uses the $13$ original clinical features. The ``PCA Quantum'' block reports the performance obtained using only the $13$ principal components fitted on the quantum features without leakage within the nested validation protocol. The ``Classical+PCA Quantum'' block reports the performance obtained from the union of the $13$ classical features and the $13$ quantum PCA components. Bold entries indicate, for each row and metric, the best quantum or mixed result whenever it exceeds the corresponding classical baseline.}
\label{tab:heart_disease_pca_quantum_mixed}
\end{table*}

\subsubsection{Sample-ordering control and hardware-drift effects}
It is important to emphasize that our results are more conservative when compared with the much larger improvements reported for the original CD-Ising-glass/DQFE pipeline on the same Molecular Toxicity dataset~\cite{simen2025digitized,simen2025quenched}. However, there is an important subtlety in hardware-executed quantum feature extraction. When samples are submitted sequentially to a QPU, the extracted features are not only functions of the input data and circuit design, but may also be affected by time-dependent calibration drift and hardware noise. If the dataset is ordered by class, as is the case for the original Molecular Toxicity file, the first and last portions of the dataset are processed at different times during the QPU run. Because the effective noise profile can evolve over time, samples appearing far apart in the submission order may pass through different noisy quantum maps, even when the nominal circuit is the same. When the submission order is also class-ordered, these time-dependent changes become aligned with the class structure. In this case, samples from one class are predominantly processed under one hardware-noise regime, while samples from the other class are processed under a later, potentially different regime. In that situation, the extracted quantum features may partially reflect acquisition order and hardware drift, rather than only input-dependent structure, thereby producing a spurious separation between the target classes. This can inflate downstream classification metrics, especially when the generated quantum features are evaluated without controlling for sample ordering.

\begin{figure}
\centering
  \includegraphics[width=\columnwidth]{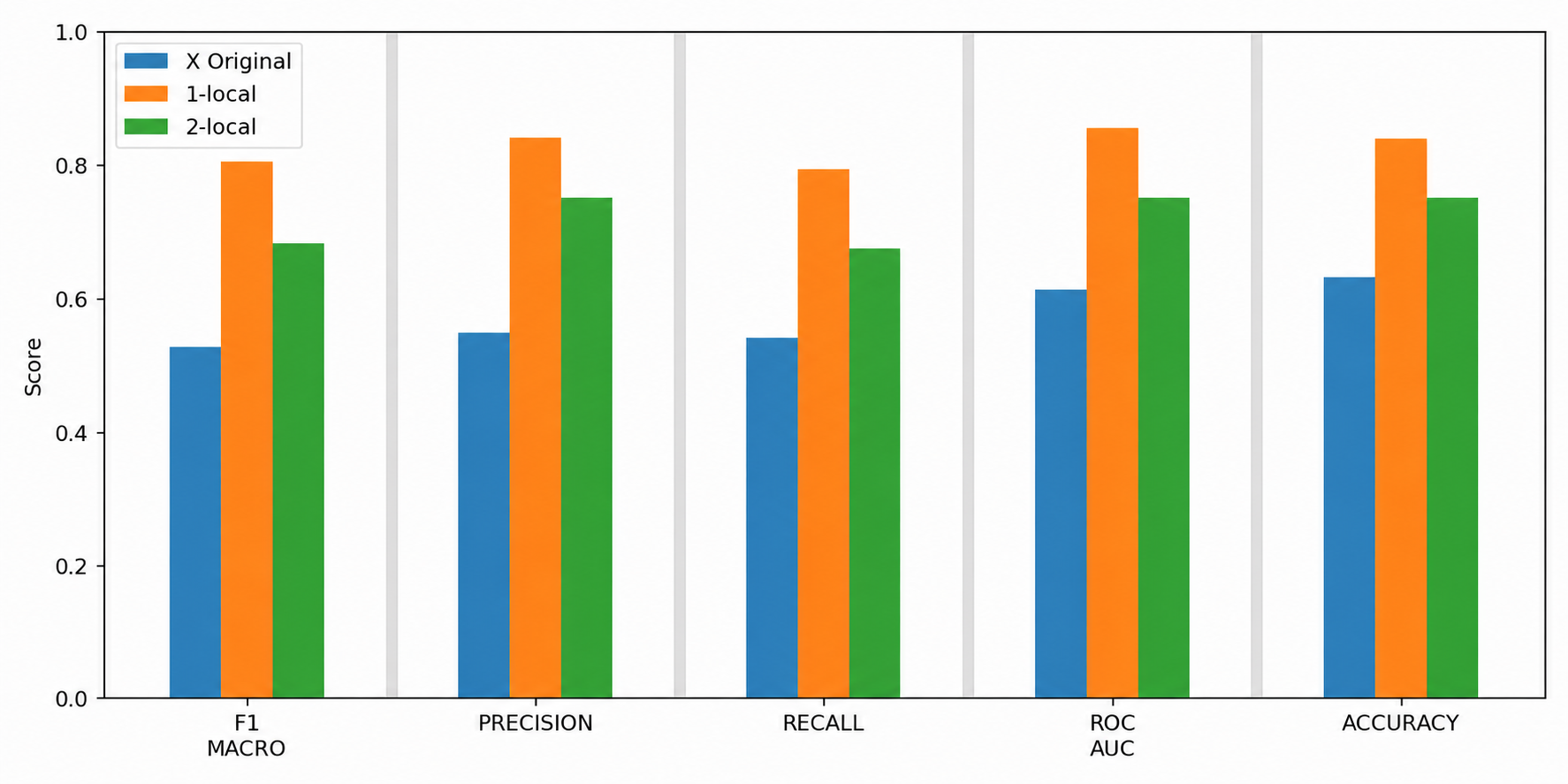}
    \caption{\justifying\label{fig:toxicity_drift}
    Performance of the CD-Ising-glass PQFM on the Molecular Toxicity dataset using 156 qubits on \texttt{ibm\_marrakesh}, without shuffling the dataset. This experiment was designed to reproduce the evaluation setting of Ref.~\cite{simen2025digitized}. It is a Gradient Boosting classifier with $1000$ estimators, random seed $42$, and stratified cross-validation with five splits and five repetitions.}
\end{figure}

We also verified this effect empirically by running the Molecular Toxicity dataset in its original class-ordered form (see Fig. \ref{fig:toxicity_drift}). In that setting, the observed gains became comparable to those reported in Refs.~\cite{simen2025digitized,simen2025quenched}, supporting the interpretation that part of the large apparent improvement can arise from an order-dependent hardware-drift effect. For this reason, the results reported here are based on shuffled samples and should be interpreted as a more conservative estimate of the predictive contribution of the PQFM itself. This makes the comparison less favorable to large apparent hardware-induced gains, but more faithful to the intended role of PQFMs as input-dependent feature maps. Similar order-dependent effects cannot be assessed for non-public datasets such as the institutional bond-trading data used in Ref.~\cite{ciceri2025enhanced}; nevertheless, the possibility of drift-induced feature separation should be considered whenever large gains are observed in long sequential QPU executions.

\subsection{High-dimensional hardware benchmark on Breast Cancer Radiology}

\begin{figure*}[t]
  \centering
  \includegraphics[width=\textwidth]{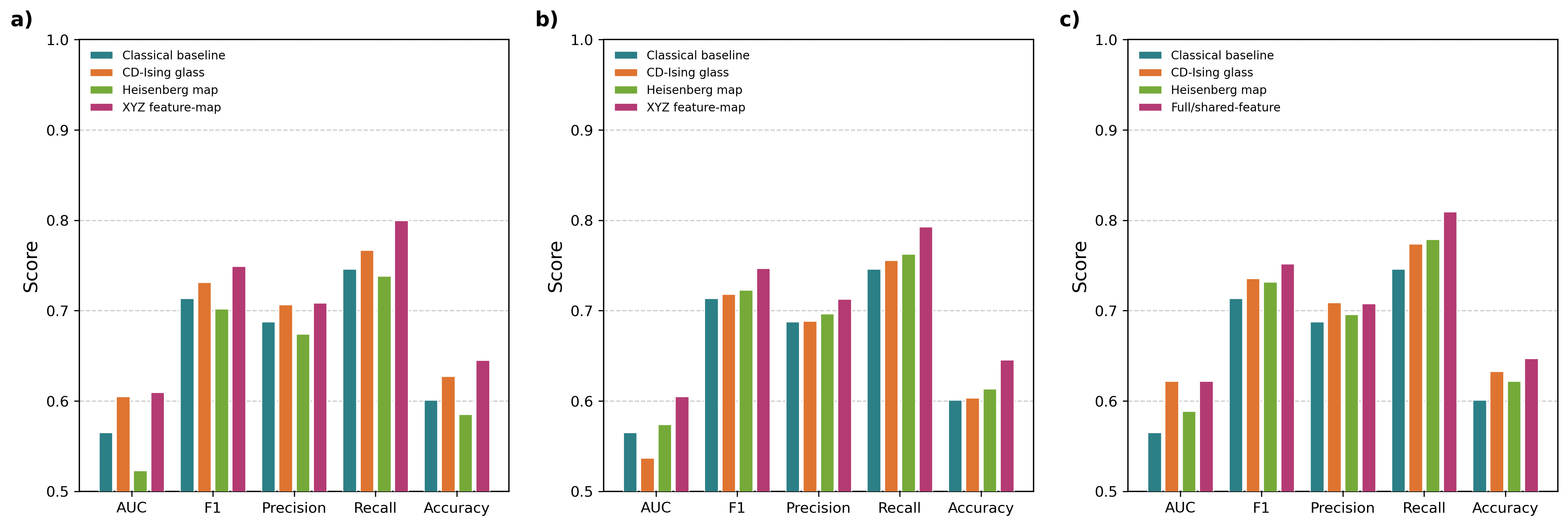}
  \caption{\label{fig:result_tox}
  \justifying Performance summary on the Molecular Toxicity dataset using the nested validation protocol shown in Fig.~\ref{fig:nested_cv_pipeline}. Bars report mean performance for AUC, F1-score, precision, recall, and accuracy; medians and standard deviations are not shown. Panels (a), (b), and (c) correspond to encoding regimes with approximately $52$, $78$, and $156$ encoded qubits, respectively. For each PQFM family and qubit budget, the displayed configuration is the one selected according to the highest mean AUC in the corresponding appendix tables in Sec.~\ref{app:toxicity_resuts}; all other metrics are taken from the same selected configuration and are not mixed across observables or settings. The CD-Ising-glass and two-qubit generalized Hamiltonian-based PQFM results are taken from \texttt{ibm\_kingston}, whereas the Heisenberg-map results are taken from \texttt{ibm\_marrakesh}. For the two-qubit generalized Hamiltonian-based PQFM family, the selected variants are XYZ-cross-only for panel (a), XY-cross-only for panel (b), and XYZ-full/shared-feature for panel (c).}
\end{figure*}

\begin{figure}[t]
    \centering
    \includegraphics[width=\columnwidth]{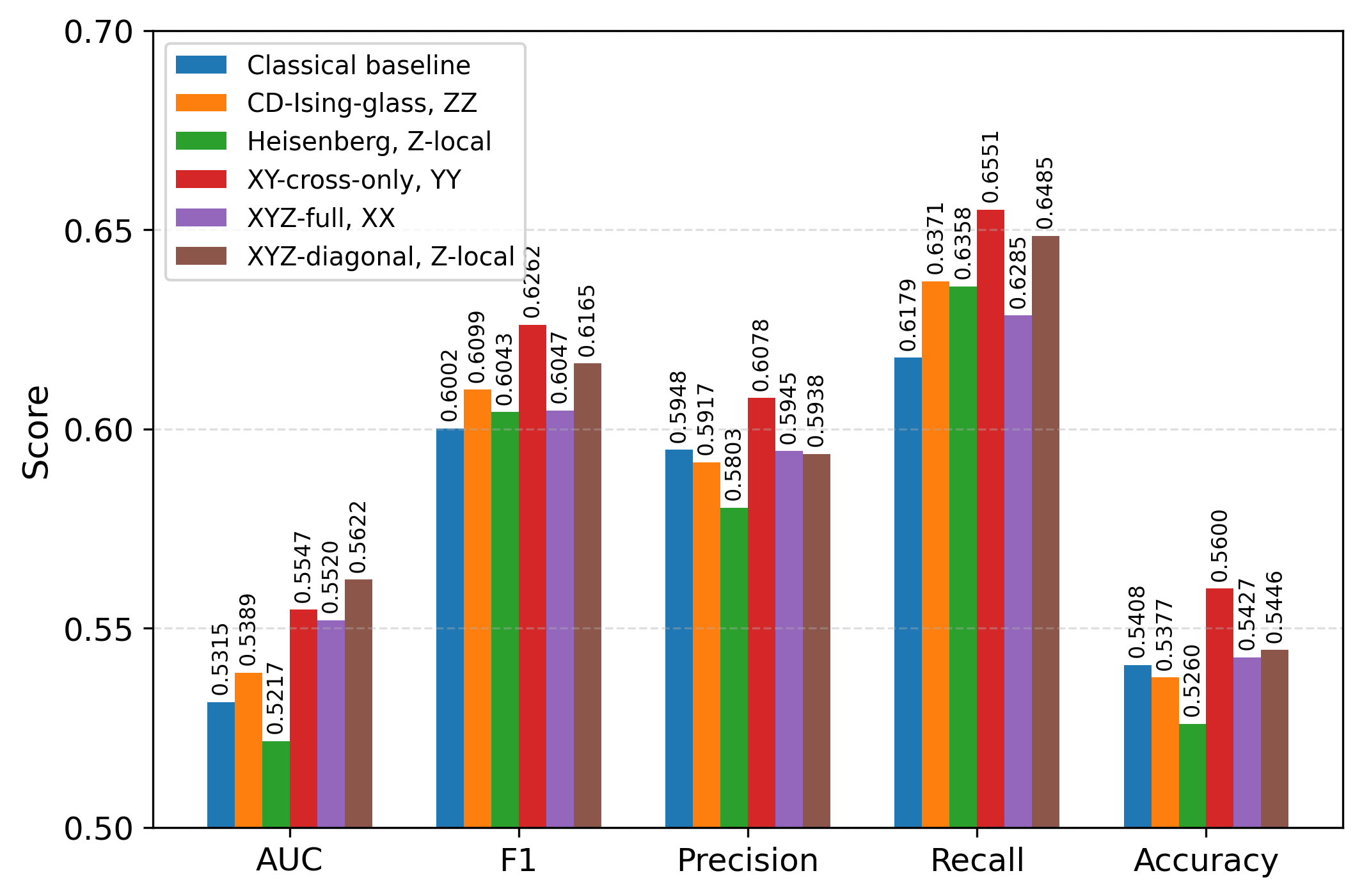}
    \caption{\justifying
    ISPY1 performance comparison on \texttt{ibm\_kingston} using $370$ input features and $161$ samples under the nested validation protocol shown in Fig.~\ref{fig:nested_cv_pipeline}. 
    Bars compare the classical baseline with the best quantum-feature configuration for each encoding family selected by AUC. 
    We consider a CD-Ising-glass encoding using $124$ qubits with $ZZ$ observables, a Heisenberg encoding using $125$ qubits with $Z$-local observables, and three multi-axis encodings: XY-cross-only using $156$ qubits with $YY$ observables, XYZ-full using $124$ qubits with $XX$ observables, and XYZ-diagonal using $124$ qubits with $Z$-local observables. 
    All configurations encode the same $370$ input features, but their qubit requirements differ due to the structure of the corresponding encoding and observable families. Among the configurations considered here, XY-cross-only is the only PQFM to use multi-feature encoding combining multi-axis and multilayer frameworks.
    }
    \label{fig:ispy1_kingston_algorithm_comparison}
\end{figure}

To further investigate the multi-feature encoding strategies and benchmark the different PQFMs, we considered the ISPY1 dataset. Unlike the Molecular Toxicity experiment, where the input space was reduced to the top \(156\) classical features, here we used all \(370\) available features. This allowed us to evaluate whether the multi-feature encoding strategies could provide an advantage in a more challenging, higher-dimensional setting. As shown in Fig~\ref{fig:ispy1_kingston_algorithm_comparison}, the generalized two-qubit Hamiltonian family again outperforms the CD-Ising-glass and Heisenberg maps, with the diagonal Hamiltonian achieving the best overall AUC. In particular, the $XYZ$-diagonal map increases AUC from 0.5315 to 0.5622 and recall from 0.6179 to 0.6485. However, the paired statistical analysis did not identify consistently significant gains for any PQFM in this experiment; only promising and suggestive gains were observed for some metrics and configurations, as reported in Table~\ref{tab:ispy1_kingston_supported_and_promising_gains}.

\subsection{Hybrid classical--quantum feature spaces in statevector simulations}

For the last two datasets, we evaluated the PQFMs in simulation and also considered hybrid classical-quantum feature spaces selected through SHAP or compressed through PCA. In contrast with the QPU experiments, the gains in the simulated setting were generally smaller. This is partly expected for the Breast Cancer Pathology dataset, where the classical baseline was already very strong, leaving little room for improvement. In this case, the Heisenberg map provided the best and statistically most consistent result among the simulated PQFMs. The SHAP-selected quantum features reached an AUC of 0.9911, compared with 0.9847 for the classical baseline, and also improved precision from 0.9574 to 0.9637 (Tables \ref{tab:breast_cancer_shap_quantum_mixed} and \ref{tab:breast_cancer_shap_supported_and_promising_gains}). The CD-Ising-glass and XYZ variants produced only small improvements when combined with classical features, while their quantum-only features were less competitive in this dataset.

Regarding the Cardiac Disease dataset, the improvements were also modest and more metric-dependent. The best gains appeared mainly in hybrid classical-quantum settings, especially when PCA was applied to the quantum features before concatenation with the classical variables. For example, the full/shared-feature $XYZ$ map improved recall from 0.7626 to 0.7900 in the classical-plus-quantum PCA setting, while CD-Ising-glass improved precision from 0.8033 to 0.8305 (Tables \ref{tab:heart_disease_pca_quantum_mixed} and \ref{tab:heart_disease_pca_supported_and_promising_gains}). However, the overall pattern was less consistent than in the hardware experiments. This may be related to the small number of input variables in this dataset. Indeed, with only 13 classical features, the Hilbert-space embedding has less opportunity to generate a substantially richer representation, and therefore the quantum feature space adds less complementary information to the downstream classifier.

\section{Conclusion}
\label{sec:conclusion}
In this work, we introduced and evaluated a generalized two-qubit Hamiltonian family of PQFMs designed to support flexible feature encoding through different Pauli-axis structures. This construction enables not only multi-axis multi-feature encoding, where each qubit can encode up to three features per encoding block, but also combinations with additional encoding blocks and shared-feature strategies, in which the same classical feature is encoded along multiple axes. We implemented this family, together with the CD-Ising-glass and Heisenberg reference maps, in the publicly available hardware-aware Python library \texttt{pqfmlib}. Across four biomedical classification datasets, including two real-QPU experiments and two simulation studies, we compared classical, quantum-only, and, when applicable, hybrid classical-quantum feature representations using a nested cross-validation protocol and paired statistical tests. Overall, our results show that Hamiltonian-based PQFMs can provide useful predictive information beyond the original classical feature space, with the most consistent gains coming from the generalized two-qubit Hamiltonian family.

In particular, in the benchmarks considered here, the multi-axis encoding strategy outperformed the alternative strategy of increasing the number of encoding blocks alone to represent more features per qubit. This suggests that distributing information across different Pauli axes is a more effective way to increase the feature capacity of near-term quantum hardware than simply adding repeated encoding layers, at least for the cases studied here. At the same time, the shared-feature mode also produced competitive results in some settings, indicating that re-encoding the same feature through different Pauli axes can expose useful complementary structure. Nevertheless, when enough qubits are available, the results indicate that the one-feature-per-qubit regime remains highly competitive, and often preferable. This raises the interesting question of whether using more qubits than classical features, for instance by re-encoding the same variables in different Hamiltonian terms or interaction patterns, could further improve performance by exposing richer quantum correlations, especially under restricted hardware connectivity.

Taken together, these results do not yet establish quantum advantage, but they provide empirical evidence that projected quantum feature maps can generate task-relevant information for classical machine-learning models. In this sense, they represent a step toward quantum utility in near-term supervised learning, especially in regimes where hardware-aware Hamiltonian design and feature-encoding strategies are jointly optimized. Moreover, the datasets used in this manuscript represent real-world problems that have a potential positive impact in medical practice. Relevant problems related to cardiovascular disease and oncology (the two main leading causes of mortality in several populations) may benefit from the application of our proposed technique. Future studies in different healthcare domains are necessary to further understand PQFM role in supporting medical practice.

However, several directions remain open. A central one is to better understand the role of hardware noise in PQFMs, including how to distinguish detrimental noise, drift-induced artifacts, and possible noise-assisted effects in the extracted feature representations. Other natural extensions include implementing Heisenberg-type maps with multi-feature multi-axis encoding, as recently proposed in related work~\cite{ciceri2025enhanced}, and generalizing the present two-qubit Hamiltonian construction to higher-order \(n\)-qubit interactions involving all three Pauli axes. These extensions may help clarify the regimes in which projected quantum feature maps can provide practical advantages for machine-learning tasks.

Finally, an important direction for making PQFMs practically scalable is the development of classical surrogate models for quantum feature extraction. Recent work by Flores-Garrigós \textit{et al.}~\cite{floresgarrigos2026offline} showed that quantum features from CD-Ising-glass can be generated only for a representative subset of the data and then learned by a classical surrogate, allowing the quantum-induced representation to be deployed at classical inference cost. Extending this idea to the PQFMs studied here would be a natural next step for large datasets where per-sample QPU evaluation is prohibitive.  At the same time, surrogate models provide a useful diagnostic benchmark, since efficient classical prediction of quantum-generated features may weaken claims of computational quantum advantage even when the resulting representation remains practically useful~\cite{huang2021power}.

\begin{acknowledgments}
The authors thank the Kipu Quantum team, especially Anton Simen, Francisco Albarrán-Arriagada, and Enrique Solano for insightful discussions and support throughout this work. This work was funded by the São Paulo Research Foundation (FAPESP) through Grant Nos.~2026/00378-3 (R. S. C.) and 2025/15490-0 (F. F. F.; QuantaNet -- FAPESP Thematic Project Grant). F. F. F. also acknowledges partial financial support from the Conselho Nacional de Desenvolvimento Científico e Tecnológico (CNPq) through the National Institute of Science and Technology for Applied Quantum Computing, Grant No.~408884/2024-0.
\end{acknowledgments}

\FloatBarrier
\appendix

\section{Molecular Toxicity results}
\label{app:toxicity_resuts}

We collect here the full Molecular Toxicity benchmark underlying the summary results discussed in the main text. The experiments compare the classical baseline with the considered PQFM families on three IBM QPUs, namely \texttt{ibm\_fez}, \texttt{ibm\_kingston}, and \texttt{ibm\_marrakesh}, under three encoding regimes with approximately $52$, $78$, and $156$ encoded qubits. The Heisenberg-map configurations use $53$ and $79$ qubits in the two smaller regimes because this construction encodes features through interaction terms.

All results follow the nested cross-validation protocol described in Fig.~\ref{fig:nested_cv_pipeline}. The outer loop uses stratified $5$-fold cross-validation with $10$ repetitions, yielding $50$ external evaluations, while hyperparameters are selected inside each outer training set using an inner stratified $80/20$ holdout split implemented with \texttt{PredefinedSplit} inside \texttt{GridSearchCV}. The reported metrics are the mean AUC, F1-score, precision, recall, and accuracy over the external evaluations.

Tables~\ref{tab:cdising_toxicity_means_only}--\ref{tab:toxicity_xyz_kingston_cross_diag_shared} summarize the results for the CD-Ising-glass, the Heisenberg-map, and the two-qubit generalized Hamiltonian-based PQFM family, respectively. For configurations involving multiple observables or feature blocks, the selected result is determined by the highest mean AUC; the remaining metrics in the same row are always taken from that same selected configuration, so metrics from different observables or settings are not mixed. Bold entries indicate quantum results that exceed the corresponding classical baseline.

\begin{table*}[t]
\centering
\scriptsize
\setlength{\tabcolsep}{3.5pt}
\renewcommand{\arraystretch}{1.15}
\resizebox{\textwidth}{!}{%
\begin{tabular}{llccccccccccccccc}
\toprule
\textbf{QPU} & \textbf{Qubits} & \multicolumn{5}{c}{\textbf{Classical}} & \multicolumn{5}{c}{\textbf{Quantum one-body}} & \multicolumn{5}{c}{\textbf{Quantum two-body}} \\
\\
\cmidrule(lr){3-7} \cmidrule(lr){8-12} \cmidrule(lr){13-17}
 & & \textbf{AUC} & \textbf{F1} & \textbf{Prec.} & \textbf{Rec.} & \textbf{Acc.} & \textbf{AUC} & \textbf{F1} & \textbf{Prec.} & \textbf{Rec.} & \textbf{Acc.} & \textbf{AUC} & \textbf{F1} & \textbf{Prec.} & \textbf{Rec.} & \textbf{Acc.} \\
\midrule
-- & -- & 0.5653 & 0.7135 & 0.6878 & 0.7461 & 0.6011 & -- & -- & -- & -- & -- & -- & -- & -- & -- & -- \\
\midrule

ibm\_fez & 52 & -- & -- & -- & -- & -- & 0.5579 & \textbf{0.7234} & \textbf{0.6899} & \textbf{0.7678} & \textbf{0.6114} & 0.4907 & 0.7044 & 0.6740 & 0.7496 & 0.5836 \\
ibm\_kingston & 52 & -- & -- & -- & -- & -- & 0.5242 & 0.7122 & 0.6818 & 0.7513 & 0.5948 & \textbf{0.6052} & \textbf{0.7314} & \textbf{0.7066} & \textbf{0.7670} & \textbf{0.6275} \\
ibm\_marrakesh & 52 & -- & -- & -- & -- & -- & \textbf{0.5693} & \textbf{0.7195} & 0.6871 & \textbf{0.7600} & \textbf{0.6047} & 0.5480 & 0.7156 & 0.6877 & 0.7548 & 0.6043 \\

\midrule

ibm\_fez & 78 & -- & -- & -- & -- & -- & 0.5206 & 0.7008 & 0.6727 & 0.7391 & 0.5823 & 0.4956 & 0.6977 & 0.6704 & 0.7383 & 0.5795 \\
ibm\_kingston & 78 & -- & -- & -- & -- & -- & 0.5371 & \textbf{0.7183} & \textbf{0.6884} & \textbf{0.7557} & \textbf{0.6035} & 0.5081 & 0.6955 & 0.6730 & 0.7304 & 0.5789 \\
ibm\_marrakesh & 78 & -- & -- & -- & -- & -- & 0.5635 & 0.7054 & 0.6734 & 0.7470 & 0.5866 & 0.5177 & \textbf{0.7198} & 0.6786 & \textbf{0.7748} & \textbf{0.6029} \\

\midrule

ibm\_fez & 156 & -- & -- & -- & -- & -- & \textbf{0.6040} & \textbf{0.7315} & \textbf{0.6992} & \textbf{0.7730} & \textbf{0.6241} & 0.5857 & 0.7287 & 0.6956 & \textbf{0.7730} & 0.6189 \\
ibm\_kingston & 156 & -- & -- & -- & -- & -- & \textbf{0.6220} & 0.7358 & \textbf{0.7092} & 0.7739 & 0.6329 & 0.6144 & \textbf{0.7480} & 0.7088 & \textbf{0.7991} & \textbf{0.6437} \\
ibm\_marrakesh & 156 & -- & -- & -- & -- & -- & \textbf{0.5956} & \textbf{0.7266} & \textbf{0.6946} & 0.7670 & \textbf{0.6153} & 0.5333 & 0.7191 & 0.6792 & \textbf{0.7713} & 0.6001 \\

\bottomrule
\end{tabular}%
}
\caption{\justifying Performance summary for the CD-Ising-glass feature map on the Molecular Toxicity dataset using the nested validation protocol shown in Fig.~\ref{fig:nested_cv_pipeline}. Results are shown for three IBM QPUs and multiple encoding regimes with $52$, $78$, and $156$ qubits. Quantum features are evaluated as one-body and two-body observables. Bold entries indicate, for each row and metric, the best quantum result whenever it exceeds the corresponding classical baseline.}
\label{tab:cdising_toxicity_means_only}
\end{table*}

\begin{table*}[t]
\centering
\scriptsize
\setlength{\tabcolsep}{3.5pt}
\renewcommand{\arraystretch}{1.15}
\resizebox{\textwidth}{!}{%
\begin{tabular}{lllcccccccccc}
\toprule
\textbf{QPU} & \textbf{Qubits} & \textbf{Features} &
\multicolumn{5}{c}{\textbf{Classical}} &
\multicolumn{5}{c}{\textbf{Heisenberg Best one-body}} \\
\\
\cmidrule(lr){4-8} \cmidrule(lr){9-13}
 &  &  &
\textbf{AUC} & \textbf{F1} & \textbf{Prec.} & \textbf{Rec.} & \textbf{Acc.} &
\textbf{AUC} & \textbf{F1} & \textbf{Prec.} & \textbf{Rec.} & \textbf{Acc.} \\
\midrule

% (52 qubits, 156 input features)
ibm\_fez & 53 & 156 &
0.5653 & 0.7135 & 0.6878 & 0.7461 & 0.6011 &
0.5444 & \textbf{0.7159} & 0.6838 & \textbf{0.7574} & \textbf{0.6036} \\

ibm\_kingston & 53 & 156 &
0.5653 & 0.7135 & 0.6878 & 0.7461 & 0.6011 &
0.5410 & 0.7025 & 0.6855 & 0.7287 & 0.5913 \\

ibm\_marrakesh & 53 & 156 &
0.5653 & 0.7135 & 0.6878 & 0.7461 & 0.6011 &
0.5233 & 0.7021 & 0.6743 & 0.7383 & 0.5855 \\

\midrule

% (78 qubits, 156 input features)
ibm\_fez & 79 & 156 &
0.5653 & 0.7135 & 0.6878 & 0.7461 & 0.6011 &
0.5241 & 0.7062 & 0.6668 & \textbf{0.7565} & 0.5831 \\

ibm\_kingston & 79 & 156 &
0.5653 & 0.7135 & 0.6878 & 0.7461 & 0.6011 &
\textbf{0.5777} & 0.7110 & \textbf{0.6951} & 0.7348 & \textbf{0.6036} \\

ibm\_marrakesh & 79 & 156 &
0.5653 & 0.7135 & 0.6878 & 0.7461 & 0.6011 &
\textbf{0.5742} & \textbf{0.7230} & \textbf{0.6966} & \textbf{0.7626} & \textbf{0.6134} \\

\midrule

% (155 qubits, 156 input features)
ibm\_fez & 156 & 155 &
0.5634 & 0.7049 & 0.6836 & 0.7330 & 0.5918 &
\textbf{0.5880} & \textbf{0.7215} & \textbf{0.6906} & \textbf{0.7609} & \textbf{0.6112} \\

ibm\_kingston & 156 & 155 &
0.5634 & 0.7049 & 0.6836 & 0.7330 & 0.5918 &
\textbf{0.5685} & \textbf{0.7113} & \textbf{0.6875} & \textbf{0.7452} & \textbf{0.6011} \\

ibm\_marrakesh & 156 & 155 &
0.5634 & 0.7049 & 0.6836 & 0.7330 & 0.5918 &
\textbf{0.5890} & \textbf{0.7319} & \textbf{0.6958} & \textbf{0.7791} & \textbf{0.6221} \\

\bottomrule
\end{tabular}%
}
\caption{\justifying Heisenberg-map results on the Molecular Toxicity dataset using the nested validation protocol shown in Fig.~\ref{fig:nested_cv_pipeline}. For each QPU and qubit budget, we report the classical baseline and the best Heisenberg one-body configuration selected according to the highest AUC among the $X$, $Y$, and $Z$ observables. All reported quantum metrics in each row correspond to the same selected observable; metrics from different observables are not mixed. Bold entries indicate Heisenberg results that exceed the corresponding classical baseline.}
\label{tab:heisenberg_toxicity_best_xyz}
\end{table*}

\begin{table*}[t]
\centering
\scriptsize
\setlength{\tabcolsep}{3.2pt}
\renewcommand{\arraystretch}{1.15}
\resizebox{\textwidth}{!}{%
\begin{tabular}{llllccccccccccccccc}
\toprule
\textbf{QPU} & \textbf{Qubits} & \textbf{Setting} & \textbf{Features} &
\multicolumn{5}{c}{\textbf{Classical}} &
\multicolumn{5}{c}{\textbf{Quantum one-body}} &
\multicolumn{5}{c}{\textbf{Quantum two-body}} \\
\\
\cmidrule(lr){5-9} \cmidrule(lr){10-14} \cmidrule(lr){15-19}
 &  &  &  &
\textbf{AUC} & \textbf{F1} & \textbf{Prec.} & \textbf{Rec.} & \textbf{Acc.} &
\textbf{AUC} & \textbf{F1} & \textbf{Prec.} & \textbf{Rec.} & \textbf{Acc.} &
\textbf{AUC} & \textbf{F1} & \textbf{Prec.} & \textbf{Rec.} & \textbf{Acc.} \\
\midrule

ibm\_kingston & 52 & cross-only, multi-axis & 156 &
0.5653 & 0.7135 & 0.6878 & 0.7461 & 0.6011 &
\textbf{0.6096} & \textbf{0.7492} & \textbf{0.7086} & \textbf{0.8000} & \textbf{0.6451} &
0.5739 & 0.7356 & 0.6956 & 0.7878 & 0.6234 \\

ibm\_kingston & 52 & diagonal, multi-axis & 156 &
0.5653 & 0.7135 & 0.6878 & 0.7461 & 0.6011 &
0.5536 & 0.7177 & 0.6997 & 0.7470 & 0.6148 &
\textbf{0.5713} & \textbf{0.7487} & \textbf{0.7113} & \textbf{0.7965} & \textbf{0.6455} \\

\midrule

ibm\_kingston & 78 & cross-only, multi-axis & 156 &
0.5653 & 0.7135 & 0.6878 & 0.7461 & 0.6011 &
\textbf{0.6049} & \textbf{0.7470} & \textbf{0.7129} & \textbf{0.7930} & \textbf{0.6456} &
0.5225 & 0.7099 & 0.6784 & 0.7539 & 0.5929 \\

ibm\_kingston & 78 & diagonal, multi-axis & 156 &
0.5653 & 0.7135 & 0.6878 & 0.7461 & 0.6011 &
\textbf{0.6088} & \textbf{0.7385} & \textbf{0.7105} & 0.7774 & \textbf{0.6359} &
0.5620 & 0.7321 & 0.6933 & \textbf{0.7843} & 0.6225 \\

\midrule

ibm\_kingston & 156 & full/shared-feature & 156 &
0.5653 & 0.7135 & 0.6878 & 0.7461 & 0.6011 &
\textbf{0.6222} & \textbf{0.7520} & \textbf{0.7079} & \textbf{0.8096} & \textbf{0.6472} &
0.5436 & 0.7067 & 0.6826 & 0.7435 & 0.5942 \\

\bottomrule
\end{tabular}%
}
\caption{\justifying Molecular Toxicity results for the XYZ feature-map family on \texttt{ibm\_kingston} using the nested validation protocol shown in Fig.~\ref{fig:nested_cv_pipeline}. The table compares cross-only, diagonal, and full/shared-feature variants under different qubit budgets. The rows with $52$ and $78$ qubits include cross-only and diagonal multi-axis settings, whereas the $156$-qubit row reports the full/shared-feature setting. For each setting, the best one-body and two-body feature blocks are selected according to the highest AUC among the available observables. For one-body features, the candidates are $X$, $Y$, and $Z$ when available; for two-body features, the candidates are $XX$, $YY$, and $ZZ$ when available. All metrics in each quantum block correspond to the same selected observable; metrics from different observables are not mixed. Bold entries indicate, for each row and metric, the best quantum result whenever it exceeds the corresponding classical baseline.}
\label{tab:toxicity_xyz_kingston_cross_diag_shared}
\end{table*}

\section{Paired Statistical Analysis of Quantum Feature Improvements}
\label{app:paired_statistical_analysis}

This appendix provides the detailed paired-delta results supporting the statistical analysis described in Sec.~\ref{sec:statistical_analysis}.Here we list the individual cases in which the selected quantum or hybrid classical--quantum feature representations improve over the corresponding classical baseline across the outer cross-validation splits.

For each dataset, algorithm, selected feature block, and metric, the tables report the mean paired delta, the corresponding 95\% confidence interval, and the one-sided Wilcoxon signed-rank $p$-value. Positive deltas indicate an improvement over the classical baseline on the same outer validation split. The status labels follow the interpretation criteria introduced in Sec.~\ref{sec:statistical_analysis}. \textit{Consistent} denotes statistically supported improvements; \textit{Promising} denotes positive deltas with Wilcoxon $p<0.05$ but confidence intervals crossing zero; \textit{Suggestive} denotes weaker positive trends with $0.05 \leq p < 0.10$ and confidence intervals crossing zero; and \textit{Positive trend} denotes positive deltas with weak statistical evidence, i.e., $p \geq 0.10$ and confidence intervals crossing zero.

The reported cases are restricted to the feature blocks selected in the corresponding performance tables, rather than to all possible observables, axes, or feature-selection strategies. Thus, this appendix should be read as a stability analysis of the best reported configurations, complementing the aggregate performance summaries in the main text. Figure~\ref{fig:stat_tox} provides a compact visualization of the statistically supported Molecular Toxicity results across the CD-Ising-glass PQFM, the Heisenberg map, and the XYZ feature-map family introduced in this work.

\begin{table*}[t]
\centering
\scriptsize
\setlength{\tabcolsep}{4.5pt}
\renewcommand{\arraystretch}{1.15}
\resizebox{\textwidth}{!}{%
\begin{tabular}{lllccccc}
\toprule
\textbf{Algorithm} & \textbf{Qubits} & \textbf{Selection} & \textbf{Metric} &
$\boldsymbol{\Delta}$ \textbf{mean} &
\textbf{IC95\%} $\boldsymbol{\Delta}$ &
\textbf{Wilcoxon greater p} &
\textbf{Status} \\
\midrule

\multicolumn{8}{l}{\textit{No selected feature block reached the Consistent criterion.}} \\

\midrule

XY-cross-only & 156 & $\langle YY \rangle$ & F1 &
+0.0260 & [-0.0022, +0.0542] & 0.0422 & Promising \\

\midrule

XY-cross-only & 156 & $\langle YY \rangle$ & Precision &
+0.0129 & [-0.0086, +0.0344] & 0.0889 & Suggestive \\

XY-cross-only & 156 & $\langle YY \rangle$ & Accuracy &
+0.0192 & [-0.0056, +0.0440] & 0.0512 & Suggestive \\

XYZ-full & 124 & $\langle XX \rangle$ & AUC &
+0.0205 & [-0.0116, +0.0526] & 0.0705 & Suggestive \\

XYZ-diagonal & 124 & $\langle Z \rangle$ & AUC &
+0.0307 & [-0.0061, +0.0675] & 0.0563 & Suggestive \\

\bottomrule
\end{tabular}%
}
\caption{\justifying
Paired statistical summary for the ISPY1 dataset, considering only the best quantum feature blocks selected in Fig~\ref{fig:ispy1_kingston_algorithm_comparison}. In this case, only results that meet the criteria of \textit{Promising}, as defined in Sec.~\ref{sec:statistical_analysis}, were found.}
\label{tab:ispy1_kingston_supported_and_promising_gains}
\end{table*}

\begin{table*}[t]
\centering
\scriptsize
\setlength{\tabcolsep}{4.5pt}
\renewcommand{\arraystretch}{1.15}
\resizebox{\textwidth}{!}{%
\begin{tabular}{lllccccc}
\toprule
\textbf{Algorithm} & \textbf{Qubits} & \textbf{Selection} & \textbf{Metric} &
$\boldsymbol{\Delta}$ \textbf{mean} &
\textbf{IC95\%} $\boldsymbol{\Delta}$ &
\textbf{Wilcoxon greater p} &
\textbf{Status} \\
\midrule

Heisenberg & 19/20 & SHAP Quantum & AUC &
+0.0064 & [+0.0008, +0.0121] & 0.0028 & Consistent \\

Heisenberg & 19/20 & SHAP Quantum & Precision &
+0.0062 & [+0.0004, +0.0121] & 0.0173 & Consistent \\

\midrule

Heisenberg & 19/20 & SHAP Quantum & F1 &
+0.0041 & [-0.0002, +0.0084] & 0.0469 & Promising \\

Heisenberg & 19/20 & SHAP Quantum & Accuracy &
+0.0053 & [-0.0002, +0.0108] & 0.0214 & Promising \\

Heisenberg & 19/20 & SHAP Classical+Quantum & Precision &
+0.0046 & [-0.0002, +0.0094] & 0.0324 & Promising \\

\midrule

Heisenberg & 19/20 & SHAP Classical+Quantum & Accuracy &
+0.0037 & [-0.0015, +0.0089] & 0.0653 & Suggestive \\

XYZ-cross/shared-feature & 20 & SHAP Classical+Quantum & Recall &
+0.0028 & [-0.0017, +0.0073] & 0.0729 & Suggestive \\

CD-Ising-glass & 20 & SHAP Classical+Quantum & Precision &
+0.0022 & [-0.0020, +0.0064] & 0.0985 & Suggestive \\

\bottomrule
\end{tabular}%
}
\caption{\justifying Statistically supported and promising improvements over the classical baseline for the Breast Cancer Pathology dataset. The table reports the statistically supported and promising cases according to the criteria defined in Sec.~\ref{sec:statistical_analysis}.}
\label{tab:breast_cancer_shap_supported_and_promising_gains}
\end{table*}

\begin{table*}[t]
\centering
\scriptsize
\setlength{\tabcolsep}{4.5pt}
\renewcommand{\arraystretch}{1.15}
\resizebox{\textwidth}{!}{%
\begin{tabular}{lllccccc}
\toprule
\textbf{Algorithm} & \textbf{Qubits} & \textbf{Selection} & \textbf{Metric} &
$\boldsymbol{\Delta}$ \textbf{mean} &
\textbf{IC95\%} $\boldsymbol{\Delta}$ &
\textbf{Wilcoxon greater p} &
\textbf{Status} \\
\midrule

CD-Ising-glass & 13 & Classical+PCA Quantum & Precision &
+0.0272 & [+0.0061, +0.0484] & 0.001776 & Consistent \\

XYZ-cross/shared-feature & 13 & Classical+PCA Quantum & Recall &
+0.0273 & [+0.0075, +0.0471] & 0.005875 & Consistent \\

\midrule

CD-Ising-glass & 13 & Classical+PCA Quantum & Accuracy &
+0.0119 & [-0.0034, +0.0272] & 0.02579 & Promising \\

\midrule

CD-Ising-glass & 13 & Classical+PCA Quantum & F1 &
+0.0096 & [-0.0078, +0.0271] & 0.08863 & Suggestive \\

Heisenberg & 13 & Classical+PCA Quantum & Recall &
+0.0158 & [-0.0012, +0.0327] & 0.05582 & Suggestive \\

XYZ-cross/shared-feature & 13 & PCA Quantum & Recall &
+0.0144 & [-0.0081, +0.0369] & 0.08833 & Suggestive \\

XYZ-cross/shared-feature & 13 & Classical+PCA Quantum & F1 &
+0.0136 & [-0.0031, +0.0302] & 0.09290 & Suggestive \\

\midrule

Heisenberg & 13 & Classical+PCA Quantum & AUC &
+0.0071 & [-0.0082, +0.0224] & 0.3497 & Positive trend \\

Heisenberg & 13 & Classical+PCA Quantum & F1 &
+0.0062 & [-0.0083, +0.0207] & 0.2947 & Positive trend \\

XYZ-cross/shared-feature & 13 & Classical+PCA Quantum & AUC &
+0.0090 & [-0.0045, +0.0225] & 0.2573 & Positive trend \\

XYZ-cross/shared-feature & 13 & Classical+PCA Quantum & Accuracy &
+0.0076 & [-0.0073, +0.0224] & 0.1908 & Positive trend \\

XYZ-cross-only/shared-feature & 13 & Classical+PCA Quantum & Recall &
+0.0129 & [-0.0055, +0.0314] & 0.1214 & Positive trend \\

\bottomrule
\end{tabular}%
}
\caption{\justifying
Paired statistical summary for the Cardiac Disease dataset using PCA on quantum features with 13 PCA components. The reported cases are grouped according to the status labels defined in Sec.~\ref{sec:statistical_analysis}: \textit{Consistent}, \textit{Promising}, \textit{Suggestive}, and \textit{Positive trend}.}
\label{tab:heart_disease_pca_supported_and_promising_gains}
\end{table*}

\clearpage
\FloatBarrier

\bibliographystyle{apsrev4-2}
\bibliography{bibl}

\end{document}